\documentclass[prd,twocolumn,tightenlines,superscriptaddress,preprintnumbers,floatfix,nofootinbib]{revtex4}

\usepackage{graphicx}
\usepackage{amsmath}
\usepackage{amssymb}
\usepackage{bbold}
\usepackage{color}
\input epsf

\newcommand{\w}{\omega}

\renewcommand{\Re}{{{\rm Re}\,}}
\renewcommand{\Im}{{{\rm Im}\,}}
\newcommand{\td}{{t_R}}
\newcommand{\taud}{{\tau_R}}

\begin{document}

\title{Analytic structure of nonhydrodynamic modes in kinetic theory}
\author{Aleksi Kurkela}
\affiliation{Theoretical Physics Department, CERN, CH-1211 Gen\`eve 23, Switzerland}
\affiliation{Faculty of Science and Technology, University of Stavanger, 4036 Stavanger, Norway}
 \author{Urs Achim Wiedemann}
\affiliation{Theoretical Physics Department, CERN, CH-1211 Gen\`eve 23, Switzerland}
\preprint{CERN-TH-2017-255}
\begin{abstract}
How physical systems approach hydrodynamic behavior is governed by the decay of nonhydrodynamic modes. 
Here, we start from a relativistic kinetic theory that encodes relaxation mechanisms governed by different timescales thus sharing essential features of generic 
weakly coupled nonequilibrium systems. By analytically solving for the retarded correlation functions, we clarify how branch cuts arise generically from noncollective
particle excitations, how they interface with poles arising from collective hydrodynamic excitations, and to what extent the appearance of poles remains at best an
ambiguous signature for the onset of fluid dynamic behavior. We observe that processes that are slower than the hydrodynamic relaxation timescale can make a system that has already reached fluid dynamic behavior to fall out of hydrodynamics at late times. In addition, the analytical control over this model allows us to explicitly demonstrate how the hydrodynamic gradient expansion
of the correlation functions can be Borel resummed such that the full nonperturbative information is recovered using perturbative input only.
\end{abstract}

\maketitle

\section{Introduction}
A broad range of physical phenomena is involved in how relativistic nonequilibrium systems reach thermal equilibrium. For near-equilibrium systems, these mechanisms are expected to leave characteristic traces in the analytic structure of the retarded correlation function of conserved quantities $G_R(\omega, k)$. On the one hand, the prototypic longtime behavior of the correlation functions that describes collective excitations evolving towards global equilibrium is given by hydrodynamic poles, whose locations and residues are dictated by the fluid dynamical gradient expansion. On the other hand the question at which time scales hydrodynamic behavior emerges, and with which confounding mechanisms it may compete, is related to the existence and properties of other nonanalytic structures in the lower complex half plane of the correlators. These nonhydrodynamic modes have been seen to govern the approach to hydrodynamics -- or hydrodynamization -- not only in static but also in rapidly evolving backgrounds, used in the phenomenological description of heavy-ion collisions \cite{Heller:2013fn,Heller:2016rtz,Denicol:2016bjh,Heller:2015dha,Casalderrey-Solana:2017zyh,Spalinski:2017mel}. While much of the recent work on nonhydrodynamic modes has focused on strongly coupled theories \cite{Starinets:2002br,Son:2002sd,Hartnoll:2005ju,Grozdanov:2016vgg},
the present study will deal with nonhydrodynamic modes in weakly coupled theories.
 
Additional motivations for studying nonhydrodynamic modes in relativistic equilibrating systems come from the apparent phenomenological need to understand how fluid dynamical behaviour arises in nucleus-nucleus, nucleus-nucleon and possibly proton-proton collisions~\cite{Khachatryan:2016txc,Aaboud:2017acw}.
Some phenomenologically successful descriptions of these systems interface hydrodynamics with transport models (see e.g.~\cite{Weller:2017tsr,Bernhard:2016tnd}), while others do not invoke hydrodynamics explicitly (see e.g. ~\cite{He:2015hfa}). This asks for a better understanding of where and how kinetic theory differs from hydrodynamics. The standard way of relating kinetic theory to viscous hydrodynamics is to derive the latter by truncating the former to a finite set of moments of the distribution function~\cite{Israel:1979wp,Denicol:2014loa}. However, this truncation is based on the assumption that hydrodynamics works. To understand whether, when, and how it breaks down necessitates investigating kinetic theory beyond the moment expansion. The purpose of the present manuscript is to do so by studying how small deviations from thermal equilibrium relax in a full kinetic theory framework.

\subsubsection{Analytic structure at strong and weak coupling} 
In known examples of strongly coupled systems at large $N_c$, the remarkable simplicity
of the microscopic structures of nonabelian plasmas is reflected in a remarkably simple analytic structure
of the full field theoretic correlation functions. 
More specifically, in $\mathcal{N}=4$ SYM theory in the limit of large number of colors $N_c \rightarrow \infty$ and
strong coupling $\lambda = g^2 N_c \rightarrow \infty$, the retarded correlation functions are known to 
exhibit an infinite set of nonhydrodynamical poles located (asymptotically for large $n$) at $\omega_{n}^{\pm} = \omega^{\pm}_0 \pm 2\pi n T (1 \mp i)$, with $n \in [1,2,3,\ldots]$, and $\omega^{\pm}_0/\pi T = \pm 1.2139 - 0.7775 i$ \cite{Starinets:2002br,Son:2002sd,Hartnoll:2005ju}. In addition, in the channels where energy momentum conservation demands, the correlation functions exhibit poles whose locations and residues are dictated for small $k$ by the hydrodynamic gradient expansion.

In weakly coupled theories, the analytic structure of retarded correlation functions is much richer. In these theories, there is a scale separation 
between the typical size of the wave packets $1/T$ and the 
mean free path between the individual scatterings $t_{\rm scat}$.  Therefore for time separations larger 
than $\Delta t \gg 1/T$, when interference effects can be neglected, the correlation function is determined by Boltzmann transport theory,
in which the collision kernels are given by in-medium scattering processes in the field theory \cite{Jeon:1994if,Jeon:1995zm,Arnold:1997gh, Arnold:2002zm}. 
The nonanalytic features of the full field theory that are absent in the transport theory are well known (see Sec.~\ref{sec2a}).
However, the nonanalytic structures appearing in the transport theory are less well understood, and will
be the topic of this contribution. As transport theory has a wider regime of validity than hydrodynamics but encompasses it, understanding these
structures provides a technically controlled in-road to understanding the onset of hydrodynamic behaviour in weakly coupled theories. 

\subsubsection{Kinetic theory in the relaxation time approximation}
While there have been numerous numerical studies of the full collision kernel in nonabelian gauge theories
\cite{Arnold:2003zc, Huot:2006ys, York:2008rr, York:2014wja, Hong:2010at, Kurkela:2014tea, Kurkela:2015qoa, Keegan:2015avk}, 
including computations of equilibrium and nonequilibrium retarded correlation functions~\cite{Keegan:2016cpi},
the question of analytical structures has been
addressed only recently~\cite{Romatschke:2015gic} in the simplest possible model of the collision kernel -- that of simple relaxation time $\tau_R$. 
In this relaxation time approximation (RTA), an ostensibly crisp and simple 
picture of the onset of fluid dynamic behaviour appears by a migration of a hydrodynamic pole through a nonhydrodynamic cut for a specific value of Knudsen number
$K = k\, \tau_R $ where $k$ is the wave number of the perturbation~\cite{Romatschke:2015gic}.
However, this simple model forgoes much of the structures of the collision kernel 
in favour of a single relaxation time.  The question of whether this simple picture survives the inclusion of more realistic collision processes is 
the starting point of this paper. 

The full weak coupling dynamics contains nonhydrodynamic excitations  at different
energy scales that relax at widely different time scales.  A minimal way of 
incorporating this generic qualitative feature while maintaining an analytically tractable model is to extend the standard RTA to a model with a 
momentum dependent relaxation time
\begin{equation}
p^\mu \partial_\mu f = \frac{p^0}{\taud(p)}(f - f_{eq})\, .
\label{eq1}
\end{equation}
For a power law form of the relaxation time 
\begin{equation}
  \taud(p)= \td (p/T)^{\xi}\, ,
  \label{eq2}
 \end{equation}  
 such a model has been used before to gain insight into freeze-out dynamics~\cite{Dusling:2009df}.

By including the scale dependence of $\tau_R(p)$, we supplement the standard
RTA approximation with features that are known to exists in QCD and other 
field theories of nonabelian plasmas. In particular, for extreme out-of-equilibrium perturbations, a.k.a. jets, the relaxation is related to the famous
jet stopping time~\cite{Baier:1996kr,Baier:1996sk}
\begin{equation}
t_{jet}(p) \sim \frac{1}{\alpha^2 T} \left(\frac{p}{T}  \right)^{1/2},
\label{eq3}
\end{equation}
corresponding to the value $\xi=1/2$ in our model. Moreover,
this generalized model shares features of bottom-up thermalization \cite{Baier:2000sb}
in the sense that decaying particles will heat up the thermal bath locally (see  
discussion at Sec.~\ref{sec3}). Both features appear generically for $\xi > 0$ while 
they are not realized in the exceptional case $\xi=0$. Other characteristic features of
QCD thermalization processes are not realized in the simple model (\ref{eq1}). 
For instance, according to (\ref{eq1}), hard particles decay directly to 
 the thermal bath while this process proceeds in full QCD via a cascade of intermediate 
 quasi-democratic splittings~\cite{Baier:2000sb,Blaizot:2013hx}.   Therefore, we cannot exclude that additional analytical
 structures of retarded correlations functions might arise in full QCD that cannot be 
 illustrated in an analysis of (\ref{eq1}). 
 However, as the analytic structures established in this manuscript for the
 model (\ref{eq1}) arise from generic features of kinetic theory, we expect 
 them to be realized in more complete descriptions, too.

\begin{figure}
\includegraphics[width=0.5\textwidth]{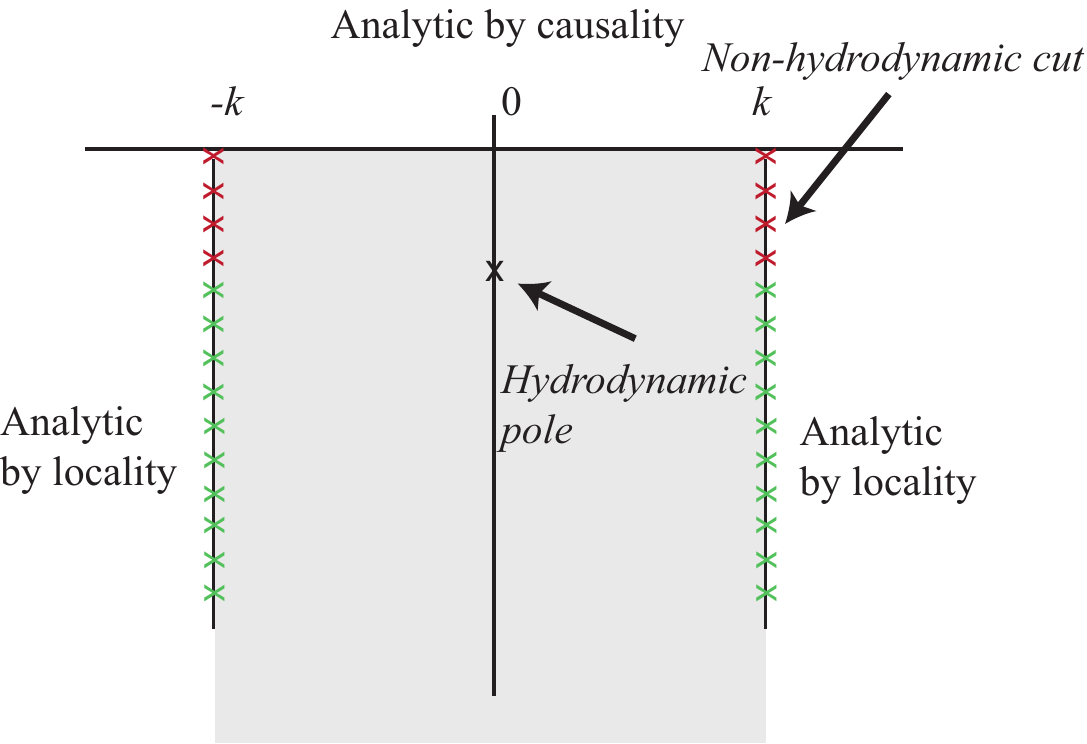}
\caption{Analytic structure of the retarded energy momentum correlation function in the shear channel $G^{0x,0x}(\omega, k)$ in the 
complex frequency plane $\omega$ for the kinetic theory (\ref{eq1}). The parts of the cut marked with red crosses correspond to medium constituent particles with lifetimes longer than the hydrodynamical decay time and will eventually dominate the correlation function at late times. The upper complex half plane is analytic by causality whereas for $|{\rm Re}\,\omega | > k$ the correlation function is analytic by locality of the scattering kernel. The nonanalytic features of the function are confined to the grey area.}
\label{fig1}
\end{figure}

The main result of the present paper is to establish the analytic structure of the retarded correlators
of the energy-momentum tensor for the model (\ref{eq1}). This result is sketched in Figure~\ref{fig1} 
for the (analytically continued) shear channel correlation function 
obtained from the model. Causality and the stability of thermal equilibrium make the correlation function analytic 
in the upper complex half-plane, while the locality of the collision kernels in the Boltzmann
equation makes the correlation function analytic for $|\Re \w| > k$. 
 In addition to the hydrodynamic pole, the model exhibits
two nonhydrodynamic cuts whose branch points are located at $\omega = \pm k$.
For any $k$, the cuts extend to smaller imaginary parts than the hydrodynamic pole; it is these structures that
are responsible for a nontrivial competition between hydrodynamics and nonhydrodynamic modes that we
discuss in detail.

The paper is organized as follows:
In section~\ref{sec2}, we first provide simple qualitative arguments for the physical mechanisms and corresponding
analytic structures arising in full gauge theories. For the class of models (\ref{eq1}), 
section~\ref{sec3} derives then explicit expressions for the retarded correlation functions. For the case $\xi = 1$,
these correlation functions can be expressed in terms of one single, analytically known generating function $H$ that
largely determines the analytic structure of the correlation functions. A detailed discussion of this analytic structure,
its physical meaning, and its ambiguities is the focus of section~\ref{sec4}, before we turn in section~\ref{sec5}
to a discussion of the physical response on pre-hydrodynamic, hydrodynamic and post-hydrodynamic time scales. 
As our study provides explicit analytic control over a model of significant physical complexity, it is also an
interesting scholarly playground for understanding how Borel resummation techniques can be applied to 
the asymptotic hydrodynamic gradient expansion. This will be discussed in section~\ref{sec6}, before we conclude
with a short summary of main results and open questions.

\section{Generic analytic properties of retarded correlators and their physical origin}
\label{sec2}

Before analyzing in detail the model (\ref{eq1}) in subsequent sections, we discuss here 
generic features of the analytic structure of retarded correlation functions of the energy momentum tensor.
In particular, we aim at providing physical intuition for the features appearing in kinetic theory. 

\subsection{Analyticity properties of retarded correlation functions in gauge theories at finite temperature}
\label{sec2a}
 
 \begin{figure}
\includegraphics[width=0.2\textwidth]{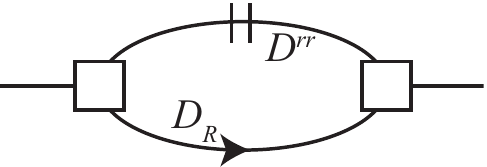} 
\caption{Diagram of (\ref{eq4}) contributing to a retarded correlation function.}
\label{diag}
 \end{figure}
 
At weak coupling the analytic structure of the retarded correlation function for $\omega \gg 1/t_{\rm scat} \sim g^4\, T$ has been 
discussed in the context of theories with different field content as 
well as in terms of different operators \cite{Arnold:1997gh,Hartnoll:2005ju,CaronHuot:2009ns,Meyer:2008gt,Laine:2011xm,Laine:2013vpa}. Quite generally, the two point
function of composite operators constructed from two field operators (such as $T^{\mu \nu}$ or the electromagnetic current $J^\mu$ of a charged field) is given to leading order by the simple one loop diagram depicted in Fig.~\ref{diag}. In the time domain, this diagram is of the generic form 
\begin{align}
G_R(t, \vec k) &\sim \int_p V(p,k) D_R(t, \vec p - \vec k )D^{rr}(t, \vec p), \nonumber \\
& \sim \int_p  \frac{-i V(p,k)}{2 E_{p-k}E_p} \theta(t)\left( e^{i E_{p-k} t}- e^{-i E_{p-k} t}\right) \nonumber \\
&\hspace{1cm}[\frac{1}{2}+n(E_p)] \left( e^{i E_p t} + e^{-i E_p t}\right)\, ,
\label{eq4}
\end{align}
where $D^R$ stands for the retarded propagator, $D^{rr} = D^> + D^<$ is the symmetric one, and $E_p = \sqrt{p^2 + m^2}$ denotes the energy associated with an excitation of momentum $p$.  
The vertices combine to a function $V$ which depends on the theory and the particular channel studied and is a function of momenta $\vec p$ and $\vec k$.  For specific cases, see \cite{Hartnoll:2005ju,Meyer:2008gt} for gauge theories. 

The
correlator (\ref{eq4}) can be decomposed naturally into two parts
\begin{align}
G_R(t,k) &\sim  C(t,k) + D(t,k)
\end{align}
that contain slowly oscillating modes of frequencies $\w= E_p - E_{p+q}$, and rapidly oscillating modes of frequencies $\w > E_p + E_{p+q}$, respectively,
\begin{align}
C(t,k) & = \theta(t)\int_p \frac{- i V(p,k)}{2 E_{p-k}E_p} n(E_p) \sin((E_p - E_{p-k})t)\, ,\\
D(t,k) & = \theta(t)\int_p \frac{- i V(p,k)}{2 E_{p-k}E_p} \left( \frac{1}{2} + n(E_p)\right) \nonumber \\
&\qquad \qquad \times \sin((E_p + E_{p-k})t)\, .
\end{align}
\subsubsection{Rapidly oscillating part $D(t,k)$}
The Fourier transform of the rapidly oscillating part
\begin{align}
D(\w, k) &\sim \int_p \frac{- i V(p,k)}{2 E_{p-k}E_p} \left( \frac{1}{2} + n(E_p)\right) \nonumber \\
& \qquad \times \left[ \frac{ E_p + E_{p-k}}{ (E_p + E_{p-k})^2 - \w^2} \right]
\label{eq9} 
\end{align}
has a cut that extends from $m + \sqrt{k^2 + m^2} < \pm \w < \infty$.  It will be a recurrent theme in this paper that the analytic structure of retarded correlation functions is ambiguous in the
sense that different analytic continuations in the complex frequency plane can account for the same physical response in the time domain. In the present case, this can be illustrated
by inserting for the massless theory the Matsubara representation 
$n(p) + \textstyle\frac{1}{2} = \sum_{n=-\infty}^\infty    \textstyle\frac{\beta p}{(2\pi n)^2 + (\beta p)^2} $ into (\ref{eq9}) and integrating over $p$.
It can be seen that by choosing a suitable analytic continuation of $D$ in the lower complex half-plane, the cuts  $m + \sqrt{k^2 + m^2} < \pm \w < \infty$ along the real axis can be exchanged into a series of cuts that are 
positioned deep in the negative imaginary region at (for $m=0$) ${\rm Im}\omega = -4 \pi n\, T$ and $-k < {\rm Re} \, \omega < k $ with $n \in [1,2,\ldots]$, see figure \ref{fig2} of Ref.~\cite{Hartnoll:2005ju}. 
As the nonanalytic structures in $D$ have a distance ${\cal O}\left(T\right)$ from the real $\w$-axis, the contribution $D$ decays on timescale $1/T$, and it is
 insignificant at late times when fluid dynamic behaviour is expected to take place.
\subsubsection{The slowly oscillating part $C(t,k)$ and kinetic theory}
As argued in \cite{Arnold:1997gh}, the slowly oscillating part $C$ arises from contributions that can be written in terms of expectation values of number operators.
This suggests that for small $k$, the physics contained in $C(t,k)$ can be captured by kinetic theory. In Fourier space, 
\begin{align}
C(\w,k) &\sim \int_p \frac{- i V(p,k)}{2 E_{p-k}E_p} n(E_p)\left[ \frac{ E_p - E_{p-k}}{ (E_p - E_{p-k})^2 - \w^2} \right]\, ,
\label{eq8}
\end{align}
the slowly oscillating nature of $C(\w,k)$ is reflected in a branch cut that extends along the real axis over the limited range $-k < \w < k$ (for all masses). For small $k$, 
this expression can be expanded to give 
\begin{align}
C(\w) \approx \int_p \frac{V(p,0)}{E_p^2} n(E_p)  \left[ \frac{1 }{i \w - i \vec v \cdot \vec k }\right]\, ,
\label{eq8b}
\end{align}
where $\vec v = \partial_{\vec p} E_p$ is the group velocity and the term in square brackets is the ballistic propagator of a free streaming point particle. We shall encounter the same branch-cut $-k < \w < k$ and the same integral (\ref{eq8b})
when we discuss the free kinetic theory in section~\ref{sec2b}.

The free theory calculation recalled here and presented, e.g., in \cite{Hartnoll:2005ju} is insufficient for $\omega \sim 1/t_{\rm scat} \sim g^4\, T$, where interactions
change the dynamics qualitatively. It therefore does not reveal the hydrodynamic pole which is close
to the origin at $\omega \sim g^4\, T$. To obtain even at leading order complete results in this region,
a class of ladder diagrams needs to be resummed~\cite{Jeon:1994if}. Such resummation can be dressed in the language
of an effective kinetic theory \cite{Jeon:1995zm,Arnold:2002zm} of nearly massless quasiparticles, where the resummed
diagrams appear in the particular scattering kernels of the kinetic equation. The effective kinetic theory is suitable for
the computation of correlation functions of the quantum field theory with external momenta $\omega, k \lesssim 1/t_{\rm scat}$, and 
therefore it is suitable for studying the vicinity of the slowly oscillating cut of $C$ in more detail than the unresummed calculation. However,
this resummation fails for larger (negative imaginary) values of $\omega$ and does not capture the physics of cuts of $D$. 

\subsection{Analytic structure of retarded correlation functions in kinetic theory}
\label{sec2b}
In this subsection, we develop an intuitive understanding for the analytic structures
accessible via kinetic theory.

\begin{figure*}
\includegraphics[width=0.4\textwidth]{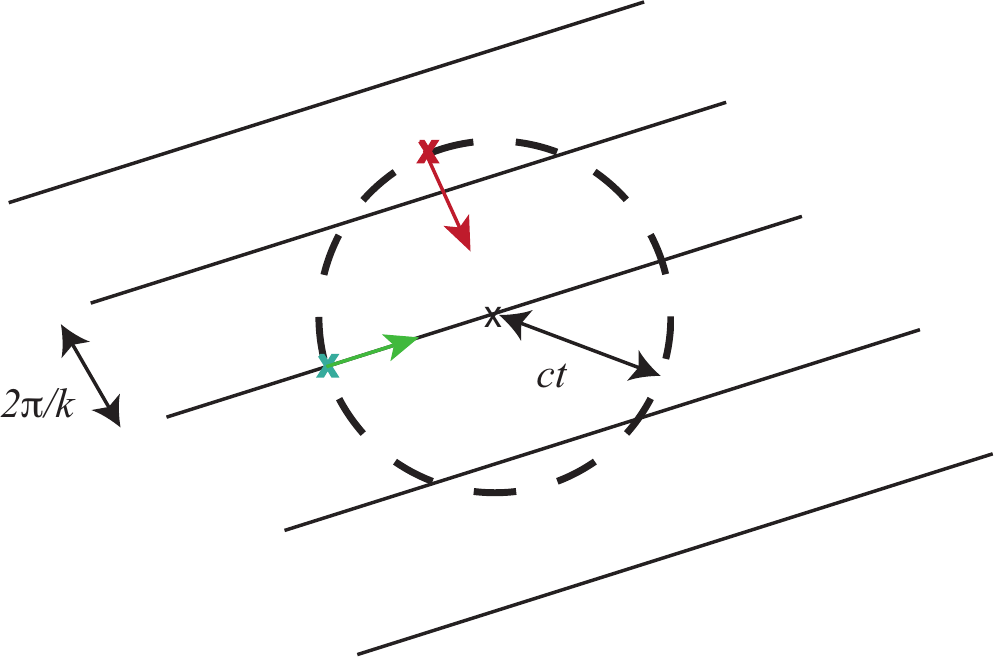}
\includegraphics[width=0.4\textwidth]{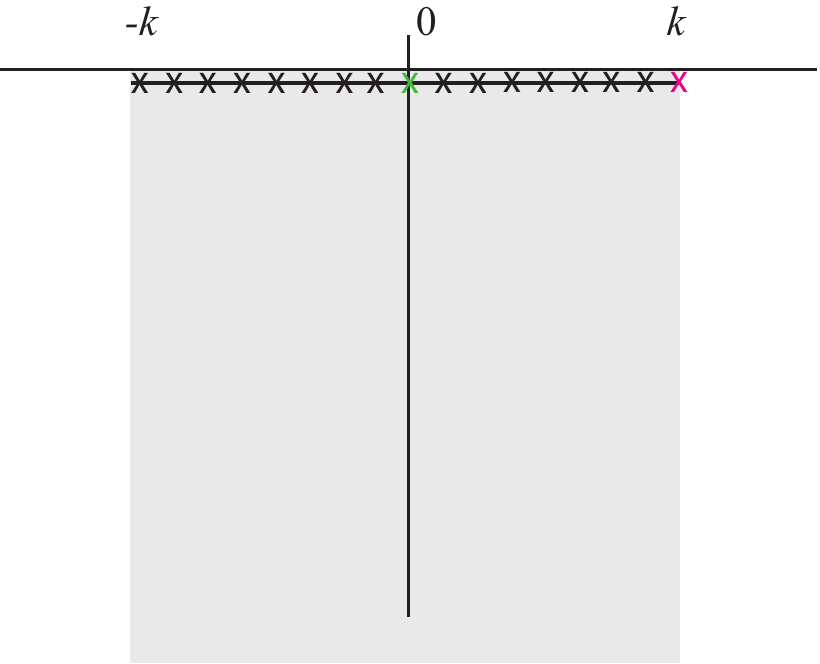}
\caption{
Left hand side: schematic picture of a perturbation in an equilibrium state that displays sheets of overdensity at wavelength $2\pi/k$. 
For a massless, free streaming gas at time $t$, the dynamical response at a position $x$ is given by integrating contributions along the circle of radius $c\, t$. 
Right hand side: Analytic structure of the retarded correlation functions $G_R^{\alpha \beta, \gamma\delta}(\w,k)$ in the complex frequency plane. 
The physics of free streaming particles is reflected in a branch cut along the real axis.
}
\label{fig2}
\end{figure*}

\subsubsection{Massless kinetic theory without interaction}
\label{sec2b1}
As sketched on the left hand side of Fig.~\ref{fig2}, a sound channel perturbation in an equilibrium system may be viewed as embedding alternating sheets of 
overdense and underdense regions that are separated in the $z$-direction by a distance $2\pi /k$. Analogous sketches can be given for perturbations in other
channels. Computing the retarded response at time
$t$ amounts then to studying  the state of the system at some 
arbitrary point $\vec x$ which initially is on the peak of the overdense region at $t=0$ when the perturbation is introduced. 

In a massless kinetic theory without interactions, particles move on straight lines at the speed of 
light. What determines the state at the point $\vec x$ at time $t$ is then the average over a sphere 
of radius $c t$. As the overdense regions are spaced $2\pi/k$ apart, the particles moving
in -$z$ direction will give rise to a signal oscillating with frequency $\w=k$. This corresponds to a pole at $k$ in the complex $\w$ plane.
Particles coming from any other direction with velocity $\vec v$ will result in an oscillating signal with smaller frequency $\w= \vec k \cdot \vec v$, 
corresponding to a pole at $\vec k \cdot \vec v$ in the complex $\w$ plane. Integrating over all orientations $\vec v$ from which particles reach 
the point $\vec x$, one finds a string of poles between $-k < \w < k$ that assemble to a logarithmic cut 
\begin{align}
\int \frac{d \Omega}{4\pi }\frac{1}{i\w - i \vec v \cdot \vec k} = \frac{i}{2 k}\log\left(\frac{\w-  k}{\w +  k}\right)\, .
\label{eq11}
\end{align}
This cut is also well known in the physics of hard thermal loops, where it gives rise to Landau damping \cite{Blaizot:2001nr}. We conclude that
the simple picture of a homogeneous and isotropic free-streaming dynamics explains the logarithmic branch cut found in interaction-free massless
kinetic theory for retarded correlation functions like, e.g., the correlation function in the sound channel calculated in~\cite{Romatschke:2015gic}
\begin{align}
G^{00,00}_R(\w,k)  = - s T  \frac{3 \w}{2 k} \log\left(\frac{\w -k}{\w + k}\right)\, .
\label{eq12}
\end{align}

\subsubsection{Massless kinetic theory in the standard RTA}
\label{sec2b2}

Romatschke~\cite{Romatschke:2015gic} has studied the effect of adding interactions to the free kinetic theory
in a simplified model of momentum-independent relaxation time approximation with collision kernel
\begin{align}
C_{RTA}[f] =  \frac{1}{\td}\left( f - f_{eq}\right)\, ,
\label{eq13}
\end{align}
where $f_{eq}$ is the local equilibrium distribution function to which the system wants to relax, 
determined by energy and momentum conservation. The inclusion of these interactions has two qualitative effects.

 First, trivially, the free particle propagator will be damped at length scales of $\Delta x \sim \td$,
shifting the cut into the negative complex plane by an amount of $-i/\td$
\begin{align}
\int \frac{d \Omega}{4\pi }\frac{1}{i \w - i \vec v \cdot \vec k - \frac{1}{\td}} = \frac{i}{2 k}\log\left(\frac{\w- k+i/\td}{\w +  k+ i/\td}\right).
\label{eq14}
\end{align}

A more subtle effect arises as a consequence of energy-momentum conservation (see eq.~(\ref{econs}) for technical details). As the energy
and momentum from the lost particles need to go somewhere, a new collective excitation
is dynamically created in channels where the conservation demands it (sound $G^{00,00}$ and shear $G^{0x,0x}$). For small $k$, the
location and residues of these poles are dictated by the hydrodynamic gradient expansion. We will call this pole in the following hydrodynamic pole. For $k \geq \pi/2 \td$, the pole crosses the cut
and enters the next Riemann sheet, thus disappearing from the physical plane. Therefore, the model has two distinct kinematic 
regimes: one where the pole is above the cut and the late time behaviour of the system is dictated by the hydrodynamic pole,
and the other where the cut dominates the dynamics at all times. This was called the hydrodynamic onset transition in \cite{Romatschke:2015gic}.

\subsubsection{Massless kinetic theory with scale-dependent RTA}

How does the analytic structure of the retarded correlator indicate that the kinetic theory of a free-streaming gas has been
supplemented with the scale-dependent relaxation dynamics of (\ref{eq1})? In close analogy to the angular integrals 
(\ref{eq11}) and (\ref{eq14}), we expect that qualitative properties of the analytic structure of retarded correlation functions 
are captured in this case by the integral
\begin{align}
\int d p f_{eq}(p)\int \frac{d \Omega}{4\pi }\frac{1}{i\w - i \vec v \cdot \vec k - \frac{1}{\taud(p)}}\, .
\end{align} 
This indicates that relaxing the assumption of a single relaxation time will  render the correlation function
nonanalytic in the entire strip $-k < {\rm Re}\, \w < k$, ${\rm Im}\, \w < 0$, where poles at different ${\rm Re } \, \w$ correspond 
to different angles of the particles, and different ${\rm Im} \, \w$ correspond to different $p$. We shall establish
this picture in an explicit calculation in section~\ref{sec4}.  It implies that the hydrodynamic pole is always embedded in the nonanalytic structure.
The existence of a clear onset transition of hydrodynamics is therefore
a spurious feature of the simple assumption of a single relaxation time in (\ref{eq13}). 
As we discuss in the next subsection, emersing the hydrodynamic pole in a nonanalytic strip results in a subtle interplay between 
hydrodynamic and nonhydrodynamic modes that can lead to a qualitatively novel phenomenon in the long-time behavior.

\begin{figure}
\includegraphics[width=0.4\textwidth]{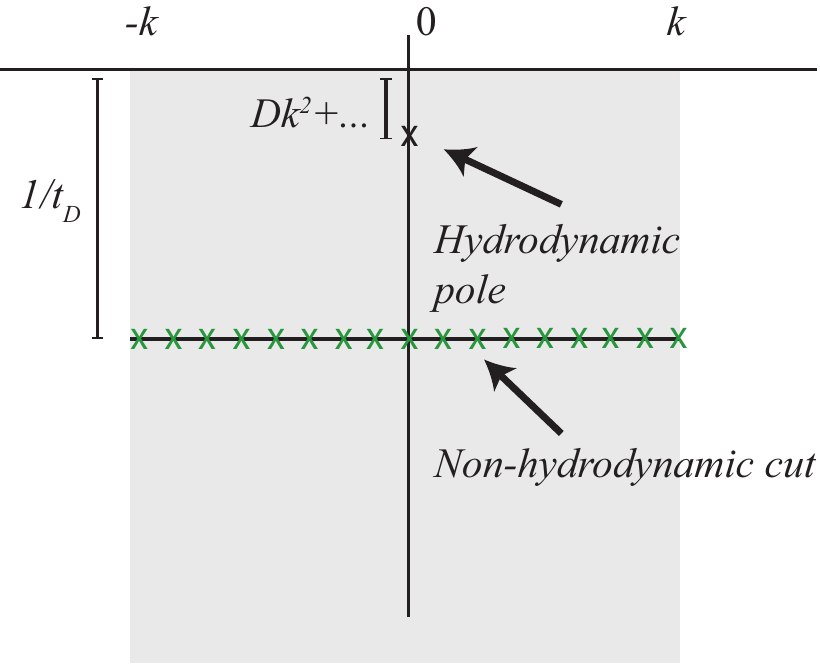}
\caption{
Analytic structure of the retarded shear correlation function $G_R^{0x, 0x}(\w,k)$ in the complex frequency plane for the kinetic theory with scale-independent relaxation time (\ref{eq13}). 
}
\label{fig3}
\end{figure}

\subsection{Dehydrodynamization in kinetic theory}
\label{sec2c}

With the simple extension to a scale-dependent relaxation time, the notion of a unique Knudsen number is obscured, as for any arbitrarily small wavenumber $k$, 
physics of different energy scales enters the transport on different time scales.   
To illustrate this parametrically, 
consider a generic small deformation of the thermal equilibrium. As by assumption the deformation does not take the system far from equilibrium, the number of perturbed modes will be, for large $p$, proportional to $e^{-\beta p}$. Each of these modes will then evolve toward equilibrium in a timescale $\tau_R(p)$, such that the overall magnitude of the nonhydrodynamic part of the perturbation 
can be estimated at time $t$ by
\begin{equation}
\delta T^{\mu \nu}(t) \sim \int_p e^{- \beta p} e^{- \frac{t}{\tau_R(p)}}\, .
\end{equation}
For a given $t$, the integral is dominated by the decay of modes at a characteristic
scale 
\begin{align}
p_*(t) \sim T \left( \xi  \frac{t}{\td} \right)^{\frac{1}{1+\xi}}\, ,
\end{align}
and the perturbation has then an overall magnitude of
\begin{equation}
\delta T^{\mu \nu} \sim e^{ - \frac{(1+\xi)}{\xi}\frac{p_*(t)}{T}}.
\label{eq17}
\end{equation}
In channels where conservation laws so demand, the deformation may also excite 
modes which relax on hydrodynamic time scales
\begin{equation}
\delta T_{hydro}^{\mu \nu} \sim e^{ - D k^2 t}\, ,
\label{eq18}
\end{equation}
where $D \sim \td $ is the appropriate diffusion coefficient in the channel in question.
For $\xi=0$, corresponding to a single relaxation time,  both contributions turn out to be exponentials and the origin of the well defined
hydrodynamization scale discussed in the previous subsection is related to the question which contribution decays faster. 
However, for general $\xi$ the situation is obviously more intricate. Amusingly, 
for $\xi > 0$, the contribution arising from the nonhydrodynamic sector is subexponential,
and dominates the signal at late times $t\gtrsim t_{\rm out}$
\begin{equation}
t_{\rm out }\sim \frac{t_{\rm scat}^{-1/\xi}}{D^\frac{1+\xi}{\xi} k^{ 2(1+\xi)/\xi}},
\end{equation}
so that one expects that at some late time a system that was hydrodynamic, will again lose its universal fluid dynamical description and be again described by specific microscopic physics related to the dynamics of the nonhydrodynamic modes.   

This dehydrodynamization mechanism will be seen at work in the model (\ref{eq1}) of scale-dependent relaxation time, where hard 
particles still decay directly to a thermal bath and hydrodynamic fluctuations of the thermal bath are ignored. 
In the full QCD collision kernel, however, the same  process proceeds via a cascade of intermediate 
quasi-democratic splittings \cite{Baier:2000sb,Blaizot:2013hx, Kurkela:2014tea,Kurkela:2015qoa}.
Also, due to the fluctuation-dissipation theorem, there are
other sources of hydrodynamic perturbations and long-time hydrodynamic tales~\cite{Arnold:1997gh,CaronHuot:2009ns,Kovtun:2011np,Akamatsu:2016llw}. Therefore, while the mechanism 
discussed here is part of full QCD, it may not dominate the late-time behavior of the full theory.

\section{The model: momentum dependent relaxation time} 
\label{sec3}
We consider a kinetic theory of the form (\ref{eq1}), coupled to an external force $F^\alpha$ 
\begin{align}
&p^\mu \partial_\mu f(\vec x, \vec p, t) + F^{\alpha} \nabla^{(p)}_{\alpha} f(\vec x, \vec p, t) \nonumber \\
 &= \frac{p^\alpha u_\alpha}{\taud(p^\alpha u_\alpha)}\big(f(\vec x, \vec p, t) - f_{eq}( T(\vec x,t), \vec u(\vec x,t) ) \big)\, .
 \label{KT}
\end{align}
The particle distribution $f(\vec x, \vec p, t)$ fulfils the massless onshell condition $p^\alpha p_\alpha = 0$,
and it is taken to be a function of spatial momenta only, such that the partial derivative $\nabla^{(p)}_0 f \equiv 0$. 
We write $p = p^0 =|\vec p |$, and our metric convention is mostly plus $\eta^{\mu \nu} = {\rm diag}(-1,1,1,1)$. 
$\taud(p^\alpha u_\alpha)$ is the momentum dependent relaxation time defined in (\ref{eq2}).
The local target equilibrium distribution function 
\begin{align}
  f_{eq} = e^{\beta p_\alpha u^{\alpha}}
  \label{boltz}
\end{align} 
depends on four macroscopic variables, the inverse temperature $\beta = 1/T$ and the flow field $\vec{u}$, with $u_a u^a = -1$ that
need  to be adjusted locally such that the time evolution conserves energy and momentum locally
in the absence of the external force  $F^\alpha=0$. According to (\ref{KT}), the condition $\partial_\mu T^{\mu \nu}=0$ implies
\begin{align}
 \int \frac{d^3 p}{(2\pi)^3} \frac{p^\nu}{p^0} \left[  \frac{p^\alpha u_\alpha}{\taud(p^\alpha u_\alpha)} (f - f_{eq}) \right]= 0\, .
 \label{econs}
\end{align}
For the case of a scale-independent relaxation time approximation when $\xi = 0$, eq.~(\ref{econs}) implies that
the target thermal system has the same local energy density as the perturbed system. In contrast, for $\xi = 1$ when
$\taud(p^\alpha u_\alpha) = \td p^\alpha u_\alpha/T$, it is the particle number density that is the same in both systems. 
For the case $\xi = 1/2$ it is something in between. Therefore, for $\xi > 0$, the evolution of the perturbed system to
the local target equilibrium will increase the energy density of the local target equilibrium system, i.e., it will heat it
up. It is in this sense that the model displays features of bottom-up thermalization for $\xi > 0$. 

\subsection{Solution for linear perturbations induced by an external source}
\label{sec3a}

The application of an external force $F^\alpha$ reshuffles energy and momentum such that, at a given point 
in space, the local target thermal distribution $f_{\rm eq}$ is no longer the global equilibrium distribution $f_{eq}^g$ but 
rather the local thermal distribution given by the local energy and momentum densities,  $f_{\rm eq} = f_{eq}^g + \delta f_{\rm eq}$. 
Here $\delta f_{\rm eq}$ accounts for the change of the target local equilibrium distribution 
due to the external force. For a Maxwell distribution -- relevant for the high-momentum particles 
that we are concentrating on -- the $\delta f_{\rm eq}$ can be written as a local perturbation of the global distribution
\begin{align}
\delta f_{\rm eq}(\vec x, \vec p, t) = p \frac{f^g_{eq}}{T}\left[ \frac{ \delta T(\vec x, t)}{T} + v_i \delta u^i(\vec x, t) \right]\, ,
\label{eq23}
\end{align}
with $\vec v \equiv \vec p / p$. 

In the presence of a small external  force $F^\alpha$, the evolution of linear perturbations $\delta f$ on top of the global thermal equilibrium $f = f_{eq}^g + \delta f$  
can be expressed by formulating eq.~(\ref{KT})  in Fourier space 
\begin{align}
\delta f = \frac{ \frac{1}{p}F^{\alpha}\nabla_{\alpha}^{(p)} f^g_{eq} - \frac{1}{\taud(p)} \delta f_{eq}}{i \w - i  \vec v \cdot k - \frac{1}{\taud(p)}}\, .
\label{df}
\end{align}
Our convention for the Fourier transform is $Q(\w, k) = \int dt d^3 k e^{i \w t - i \vec k \cdot \vec x}Q(t,\vec x)$.
In eq.(\ref{df}), we have used the relation $f-f_{\rm eq} = \delta f - \delta f_{\rm eq}$. This relation implies also that
up to linear perturbations, eq.~(\ref{econs}) translates into constraints for four particular integral moments of $\delta f_{\rm eq}$ and $\delta f$, namely
\begin{align}
&\int \frac{d^3 p}{(2\pi)^3} \frac{ p^\nu}{\taud(p)} \delta f = \int \frac{d^3 p}{(2\pi)^3} \frac{ p^\nu}{\taud(p)} \delta f_{\rm eq} 
\nonumber \\
&= \int \frac{d^3 p}{(2\pi)^3} \frac{(p^\nu)}{\taud(p)}   f^g_{eq}\left[ \frac{p \, \delta T}{T^2} + \frac{p^i  \delta u^i}{T} 
\right] \, .
\label{eq25}
\end{align}
As a consequence, both sides of eq.~(\ref{df}) depend on $\delta f$, the left hand side explicitly and the right hand side implicitly through $\delta f_{\rm eq}$. 
The rewritten condition (\ref{eq25}) for energy-momentum conservation makes this implicit dependence manifest. The task is to solve the four equations (\ref{eq25}) 
self-consistently for the four local perturbations of the target temperature $\delta T(\vec x,t)$ and target flow fields $\delta \vec u(\vec x,t)$ that define $\delta f_{\rm eq}$.
This is done by inserting (\ref{df}) into (\ref{eq25}), thus finding a closed set of four equations for the four 
variations  $\delta T$ and $\delta u^i$. The solution of this set of equations is
\begin{align}
\delta T &=  S^0 + \delta T I^{2\xi,0,0} + \delta u_z I^{2\xi,0,1} \, ,
\label{eq26}\\
\delta u^z &= 3  S^0 + 3 \delta T I^{2\xi,0,1} + 3 \delta u_z I^{2\xi,0,2}\, , \label{eq27} \\
\delta u^x & =  S^x +  \frac{\delta u_x}{2}  I^{2\xi,2,0}\, ,\label{eq28} \\ 
\delta u^y & =  S^y + \frac{\delta u_y}{2} I^{2\xi,2,0}\, , \label{eq29}
\end{align}
where the integral moments and sources are defined by
\begin{align}
I^{abc} & = \frac{-2\pi^2}{\Gamma(5-\xi)T^5 \td}\int \frac{d^3 p}{(2\pi)^3} p^2 \frac{f_{eq}^{g}(p)  (T/p)^a v_\perp^b v_z^c}{i \w - i  \vec v \cdot \vec k - \frac{T^\xi}{\td p^\xi}}\, , 
\label{eq30} \\
S^{\mu} & =  \frac{- 2\pi^2}{\Gamma(5-\xi)T^5}\int \frac{d^3 p}{(2\pi)^3} \frac{T^\xi}{p^\xi}\frac{f_{eq}^g(p) F^i v_i v^\mu}{i\w - i \vec v\cdot \vec k - \frac{T^\xi}{\td p^\xi}}\, ,
\label{eq31}
\end{align}
with $v_\perp^2 = 1 - v_z^2$. The solutions (\ref{eq26})-(\ref{eq29}) for the perturbations of the local target temperature and flow velocity fully define the 
deviation $\delta f_{\rm eq}$ of the local target equilibrium distribution from the global equilibrium distribution. This allows one to write explicit expressions 
for all terms on the right hand side of eq.~(\ref{df}). Therefore, in terms of these solutions, eq.~(\ref{df}) contains now the full microscopic 
information of the system.  

We note as an aside that the following discussion could be easily extended to the case of Bose (or Fermi) statistics, replacing (\ref{boltz}) by the corresponding
sum over exponentials
\begin{align}
	\frac{1}{e^{\beta p} - 1} = \sum_{n=1}^\infty e^{-n \beta p}\, .
\end{align}
In particular, the integral moments (\ref{eq30}) can be simply calculated for this statistics, resulting in
\begin{align}
	I^{abc}_{\rm Bose}(T) =  \sum_{n=1}^\infty I^{abc}(T/n)\, .
\end{align}

\subsection{Retarded correlation functions}
\label{sec3b}
We follow the standard procedure of sourcing the departure of the energy-momentum tensor from equilibrium,
\begin{align}
	\delta T^{\mu\nu} = \int \frac{d^3 p}{(2\pi)^3} \frac{p^\mu p^\nu}{p^0}\, \delta f
	\label{eq32}
\end{align}
by a perturbation of the metric $g_{\mu \nu} = \eta_{\mu \nu}+h_{\mu \nu}$. This amounts to applying an external force
\begin{align}
F^i v_i = - p^2 \Gamma^i_{\alpha \beta}v^\alpha v^\beta v_i\, ,
\label{eq33}
\end{align}
where the $\Gamma^i_{\alpha \beta}$ denote Christoffel symbols. The retarded correlation functions $G_R^{\mu\nu,\alpha\beta}$
define then the response of the energy momentum tensor to the metric perturbation,
\begin{align}
	\langle T^{\mu\nu}\rangle = \frac{\partial T_{\rm eq}^{\mu\nu}}{\partial h_{\alpha\beta}}\Big\vert_{h=0} h_{\alpha\beta} - \frac{1}{2} G_R^{\mu\nu,\alpha\beta} h_{\alpha\beta} \, ,
	\label{eq34}
\end{align}
and they can be evaluated  in terms of functional derivatives 
\begin{align}
 G_R^{\mu\nu,\alpha\beta} = \frac{\delta T^{\mu\nu}}{\delta h_{\alpha\beta}}\, .
\end{align}
The disturbance  $\delta T^{\mu\nu}$ of the energy momentum tensor is given explicitly in terms of equation (\ref{df}), with $\delta f_{\rm eq}$ defined
in terms of eqs.~(\ref{eq23}) and (\ref{eq26})-(\ref{eq29}). The evaluation of the functional derivative $\delta T^{\mu\nu}/\delta h_{\alpha\beta}$ is then
straightforward and one  finds
\begin{widetext}
\begin{align}
G_R^{xy,xy} &= - i \w sT \td   \frac{\Gamma(5-\xi)}{64}I^{0,4,0} \, ,
\label{tensor} \\
G_R^{0x,0x} &=  i k sT  \td  \frac{\Gamma(5-\xi)}{16} \left[    -I^{0,2,1}- \frac{3}{2} I^{\xi,2,0} \frac{ I^{\xi,2,1}}{ 1 -\frac{3}{2} I^{2\xi,2,0} }\right]\, , \label{shear}\\
G_R^{zz,zz} &=  - i \w sT \td \frac{\Gamma(5-\xi)}{8}  \left(\frac{\left(1-3 I^{2\xi,0,2}\right) (I^{\xi ,0,2})^2+6 I^{2\xi,0,1} I^{\xi ,0,3} I^{\xi ,0,2}+3 \left(1-I^{2\xi,0,0}\right) (I^{\xi ,0,3})^2}{-3 (I^{2\xi,0,1})^2-I^{2\xi,0,0}-3 \left(1-I^{2\xi,0,0}\right) I^{2\xi,0,2}+1} + I^{0 ,0,4}\right) \, . \label{sound} 
\end{align}
\end{widetext}
These retarded correlators describe the response in the spin 2 tensor channel (\ref{tensor}) induced by $h_{xy}$, in the spin 1 shear channel (\ref{shear}) induced by $h_{0x}$ (or $h_{0y}$, $h_{xz}$, $h_{yz}$)
and in the spin 0 sound channel (\ref{sound}) induced by $h_{zz}$ (or $h_{00}$, $h_{03}$, $h_{xx}$, $h_{yy}$), respectively. 
The remaining components of the correlation functions can be obtained from relations imposed by energy-momentum conservation, such as 
$\partial_\mu G^{\mu \alpha, \beta\gamma} =0$. 
For instance, $G_R^{xz,xz}(\omega,k) = \textstyle\frac{\omega}{k} G_R^{xz,x0} (\omega, k)$,
 $G_R^{0x,0x}(\omega,k) = \textstyle\frac{k}{\omega} G_R^{0x,xz} (\omega, k)$ or $G_R^{00,00}(\omega,k) = \textstyle\frac{k}{\omega} G_R^{0z,00} (\omega, k)$ , 
 $G_R^{00,00}(\omega,k) = \textstyle\frac{k^2}{\omega^2} G_R^{0z,0z} (\omega, k)$. 
We have explicitly checked (up to high orders in the gradient expansion) that the various correlation functions satisfy these nontrivial Ward identities that are not apparent in the above calculation. 
We have also checked explicitly that for the special case of a momentum-independent relaxation time, $\xi = 0$, the retarded correlation functions (\ref{tensor}), (\ref{shear}), and (\ref{sound})
reduce to the results of Ref.~\cite{Romatschke:2015gic}.

\subsection{The fluid dynamic limit of $G_R$}

Up to second order in the gradient expansion in small $\w$ and $k$, the form of retarded correlation functions is dictated by 
second order fluid dynamics, namely
\begin{align}
G_{R,hyd}^{xy,xy} &= - i \eta \w + \frac{1}{2}\left(\kappa\left(k^2+\w^2\right) + 2 \eta \tau_\pi \w^2 \right) + \ldots \, ,
\label{tensorhyd}\\
G_{R,hyd}^{0x,0x} &=  - \frac{i k^2 \eta}{\w} + \left( \frac{\eta^2 k^4}{sT \w^2} + \eta \tau_\pi k^2 \right)+ \ldots  \, , \label{shearhyd}\\
G_{R,hyd}^{zz,zz} &=  \frac{c_s^2 sT \w^2}{-c_s^2 k^2+\w^2}  -  \frac{4 i \eta \w^5}{3 \left(-c_s^2 k^2+\w^2\right)^2} + \ldots \, , \label{soundhyd} 
\end{align}
where dots indicate terms of higher power in $k$ or $\w$. These fluid dynamic expressions depend on entropy $s$, temperature $T$, sound
velocity $c_s^2$, as well as shear viscosity $\eta$, the shear viscous relaxation time $\tau_\pi$ and the second order transport coefficient $\kappa$.
To determine these fluid dynamic parameters for the kinetic theory with scale-dependent relaxation time, we want to compare the gradient
expansion of (\ref{tensor}), (\ref{shear}) and (\ref{sound}) to the hydrodynamic expressions  (\ref{tensorhyd}), (\ref{shearhyd}) and (\ref{soundhyd}). 
To this end, we expand the integrand of the integral moments (\ref{eq30}) to arbitrary order $N$ in  $\w$ and 
$k$, and we perform the $p$-integration for each term in this expansion. This leads to  
\begin{align}
I^{abc} \approx &\sum_{R=0}^{N}  (i \omega )^R  \frac{\sqrt{\pi}\Gamma (-a+R \xi +\xi +5) \Gamma \left(\frac{b}{2}+1\right)}{ 2\Gamma (5-\xi )}  \times\nonumber \\
& \left\{
\begin{array}{ll}
 \Gamma \left(\frac{c+1}{2}\right)  {\, _3\tilde{F}_2}^{\frac{c+1}{2},\frac{1-R}{2},-\frac{R}{2}}_{\frac{1}{2},\frac{1}{2} (b+c+3)}\left(\frac{k^2}{\omega ^2}\right) & \textrm{even }c \\
-\frac{ k R  }{2 \omega }  \Gamma \left(\frac{c}{2}+1\right) {\, _3\tilde{F}_2}^{\frac{c+2}{2},\frac{1-R}{2},1-\frac{R}{2}}_{\frac{3}{2},\frac{1}{2} (b+c+4)} \left( \frac{k^2}{\omega ^2}\right)& \textrm{odd }c
\end{array}
\right.
\, ,
\end{align}
where $\,_3\tilde F_2$ is the regularized generalized hypergeometric function. If one of the upper indices of the 
hypergeometric function is zero or negative integer, the sum truncates to a hypergeometric polynomial, which is the 
case here when $R \neq 0$.
 For example 
\begin{align}
I^{040} \approx &\frac{1}{\Gamma(5-\xi)}\Big(\frac{8}{15}\Gamma(5+\xi) + \frac{8 i \w }{15}\Gamma(5+2\xi) \nonumber \\
& - \frac{8 (k^2 + 7 \w^2 ) \Gamma(5+3\xi)}{105} \nonumber \\
& - \frac{8 i (3 k^2 \w + 7 \w^3)\Gamma(5+4\xi)}{105} + \ldots \Big)\, ,
\label{eq44b}
\end{align}
which, modulo prefactors, determines the gradient expansion of the tensor channel $G_R^{xy,xy}$ in (\ref{tensor}). 
We note that this is an asymptotic series.   Comparing these gradient expansions to the hydrodynamic limits, one finds
\begin{align}
c_s^2 & = 1/3\, ,\\
\eta &= \frac{\Gamma(5+\xi)}{120} sT \td \, ,\\
\tau_\pi &=  \frac{\Gamma(5+ 2 \xi)}{\Gamma(5+\xi)} \td \, ,
\label{eq44}\\
\kappa &= 0\, ,
\end{align}
see also Ref.~\cite{Dusling:2009df}.
Given that the retarded correlators (\ref{tensor}), (\ref{shear}) and (\ref{sound}) are those of a kinetic theory of massless particles, the
speed of sound takes of course the value expected for a conformal theory. The expressions for shear viscosity $\eta$, the shear viscous relaxation 
time $\tau_\pi$, and $\kappa$ are genuine kinetic theory results. As the present evaluation is based on a linearized response to perturbations, it is 
not sufficient to determine those second order transport coefficients $\lambda_1,\lambda_2,\lambda_3$ 
which depend nonlinearly on perturbations. 

Hydrodynamic poles arise as a consequence of energy momentum conservation. In the kinetic theory calculation of section~\ref{sec3b}, the structures
in the retarded correlators that arise from energy-momentum conservation are related to the term $\delta f_{\rm eq}$ on the right hand side of (\ref{df}). 
Inserting the disturbance (\ref{df}) into (\ref{eq32}) and performing the functional derivative $\delta T^{\mu\nu}(\omega,k)/\delta h_{\alpha\beta}(\omega,k)$, 
one finds that it is exactly the nontrivial denominators in (\ref{shear}) and (\ref{sound}) that arise from the terms proportional to $\delta f_{\rm eq}$. 
The hydrodynamic poles in (\ref{shear}) and (\ref{sound}) are therefore given by the
zeroes of the nontrivial denominators in these two channels that arise from energy momentum conservation.
 
To make the pole structure of the fluid dynamic limit of retarded correlation functions more explicit, one can write 
the fluid dynamic limit of the shear and sound channels as 
\begin{align}
	G_{R,hyd}^{0x,0x} &=  \frac{\eta k^2 }{i \w\left( 1 - i \tau_\pi \w\right)  - \frac{\eta}{sT} k^2} \, ,
	\label{eq46}
\end{align}
and 
\begin{align}
	G_{R,hyd}^{zz,zz} &= sT \frac{c_s^2 \w^2 - i \frac{4}{3} \frac{\eta}{sT} \w^3 }{\w^2 -c_s^2 k^2 + i \frac{4}{3} \frac{\eta}{sT} k^2 \w} \, .
	\label{eq47}
\end{align}
These expressions agree up to second order  in gradient expansion with (\ref{shearhyd}) 
and (\ref{soundhyd}), respectively. Higher orders in the gradient expansion of the full retarded correlators cannot be expected to
be reproduced correctly by (\ref{eq46}) and (\ref{eq47}). In this sense, the precise location of fluid dynamical poles is beyond the scope of 
a second order gradient expansion. We shall discuss it in section~\ref{sec5} without taking recourse to the gradient expansion.

\section{Analytic structure of the retarded correlation function in momentum dependent relaxation time approximation}
\label{sec4}
The full retarded correlation functions are defined in terms of the integral moments $I^{a,b,c}(\w,k)$ .
To study these correlation functions beyond the simple gradient expansion,
one needs to evaluate $I^{a,b,c}(\w,k)$  for nonzero $\w$ and $k$. A numerical evaluation of $I^{a,b,c}(\w,k)$ in (\ref{eq31})
is possible for arbitrary momentum dependencies of the relaxation time approximation (\ref{eq2}), i.e., for arbitrary $\xi$.
However, analytical control is advantageous for studying the analytic structure. We therefore focus in the following sections
on the case $\xi = 1$ for which explicit analytical results can be obtained. However, we expect that the qualitative features found
for the case $\xi = 1$ extend to the generic case $\xi > 0$. 

The simplification in the case $\xi = 1$ arises from the fact that all integral moments can be related explicitly to a single generating 
function 
\begin{align}
I^{a,b,c} = R_1^{a,b,c}(\bar \w, \bar k) + R_2^{a,b,c}(\bar \w, \bar k ,\partial_\rho) H(\rho,\bar\w ,\bar k) |_{\rho =1}\, ,
\label{eq48}
\end{align}
with 
\begin{align}
\bar \w & \equiv \td \w\, , \\
\bar k & \equiv k \td\, .
\end{align}
Here, $R_1^{a,b,c}$ and $R_2^{a,b,c}$ are simple rational functions of $\w$ and $k$,
and in $R_2$ the derivative $\partial_\rho$ appears only in the numerator 
of the rational function. The generating function reads 
\begin{align}
H(\bar \w,\bar k,\rho) = \frac{1}{2}\int_{-1}^1 dx \int_0^\infty dp\frac{  p e^{-\rho p}}{(\bar \w - \bar k x) p + i} \, .
\label{eq49}
\end{align}
Appendix~\ref{appb} provides details of this reduction. According to the procedures presented there, 
a symbolic computation program for algebraic reduction~\cite{Mathematica} can be employed to obtain 
explicit expressions for the rational functions $R_1^{a,b,c}$ and $R_2^{a,b,c}$ that enter all moments $I^{a,b,c}$ of interest. 

We note as an aside that we have attempted to derive expressions similar to (\ref{eq48}) for other values of $\xi$. 
For other rational values, such as $\xi = 1/2$, $\xi = 1/3$ etc, one finds typically expressions in terms of more than
one generating function, but we were not able to bring all of them into closed analytical form. 

To evaluate $H(\bar \w,\bar k,\rho)$, we start from the representation 
\begin{align}
H(\bar  \w ,\bar  k, \rho ) &= \frac{1}{2 \bar k} \int_{-\bar k}^{\bar k} dx (-\partial_\rho)G(\bar \w - x,\rho)
\label{H}
\end{align}
in terms of the function
\begin{align}
G(\bar \w,\rho) &= \int_0^\infty dp \frac{  e^{-\rho p}}{\bar \w p + i} \nonumber\\
&=  \frac{e^{\frac{i \rho}{\bar \w}}}{\bar \w}\Gamma\left(0, \frac{i \rho}{\bar \w}\right)\, ,\label{G}
\end{align}
Here, the integration contour crosses the pole when $\bar{\omega}$ takes negative imaginary values. The incomplete gamma function 
$\Gamma\left(0, \frac{i \rho}{\bar \w}\right)$ therefore has a logarithmic branch cut for negative imaginary $\bar{\omega}$. 
%
\subsection{Analytic structure of the generating function $H$}
The analytic structure of the retarded correlation functions is determined by the analytic structure of the integral moments $I^{a,b,c} $
which in turn is mainly determined by the analytic structure of the generating function $H$. We therefore discuss now the properties of
$H$ in detail. To perform the integral in (\ref{H}), we note that $\rho G(\bar \w, \rho)$
is a function of $\bar \w/\rho$ only. The derivative with respect to $\rho$ can therefore be replaced by a derivative with respect to $x$,
\begin{align}
H(\bar \w,\bar k,\rho) = -\frac{1}{2\bar k}\int_{-\bar k}^{\bar k} dx \partial_x \left[ \frac{\bar \w-x}{\rho} G(\bar \w-x,\rho)\right] 
\label{htotalderiv}\, .
\end{align}
When integrating this total derivative, one needs to note that for $\bar{\omega}$-values with real part in the range $-\bar k<{\rm Re}\,\bar \w< \bar k$,
the $x$-integration crosses between $x={\rm Re}\,\bar \w - \epsilon$ and $x={\rm Re}\,\bar \w + \epsilon$ 
the branch cut of $\Gamma\left[0,\textstyle\frac{i\rho}{\bar{\omega}-x}\right]$ for all values $\bar{\omega}$ with
$\Im (\bar{\omega}) < 0$. The corresponding discrete contribution to the integral is proportional to
\begin{align}
& \left[ \frac{\bar \omega-x}{\rho} G(\bar \omega-x,\rho)\right]\bigg\rvert_{\textrm{Re}\bar \omega - \epsilon}^{\textrm{Re}\bar \omega + \epsilon} \nonumber \\
&= \frac{e^{\rho/{\rm Im \bar \omega}}}{\rho}\left[ \Gamma\left(0,\frac{i \rho}{i {\rm Im}\bar \omega - \epsilon}\right)-\Gamma\left(0,\frac{i \rho}{i {\rm Im}\bar \omega + \epsilon}\right)\right] \nonumber \\
&= -\frac{e^{\rho/{\rm Im}\bar \omega}}{\rho}\left[ \log\left(\frac{-\rho}{(-\Im \bar \omega)}-i \epsilon\right)-\log\left(\frac{-\rho}{(-\Im \bar \omega)}+i \epsilon\right)\right]\nonumber \\
&= \frac{e^{\rho/\Im \bar \w}}{\rho}\left[ i 2 \pi \theta(- \Im \bar \w)\right]\, .
\end{align}
Integrating the total derivative in (\ref{htotalderiv}) therefore yields
\begin{align}
H(\bar \w,\bar k,\rho)=&\frac{-1}{2k}\Bigg(\left[ \frac{\bar \omega-x}{\rho} G(\bar \omega-x,\rho)\right] \bigg\rvert_{-\bar k}^{\bar k} \label{hfunc} \\
& - 2\pi i\, e^{-\rho/\Im\bar \w}\theta(- \Im \bar \w)\theta(\bar k^2 - (\Re \bar \w)^2 ) \Bigg)\, . \nonumber
\end{align}
The analytic structure of the full retarded correlation functions inherits the analytic structure of the generating function $H$  in the sense that where
the generating function is nonanalytic, so is the full correlation function.  The nonanalytic structures seen in eq.~(\ref{hfunc}) can therefore
be related to some of the nonanalytic structures sketched for the retarded correlation function in the introductory Fig.~\ref{fig1}. In
particular, in the first line of eq.~(\ref{hfunc}), the two terms $\propto \left(\bar \omega + \bar k\right) G(\bar \omega + \bar k,\rho)$  
and $\propto \left(\bar \omega - \bar k\right) G(\bar \omega - \bar k,\rho)$ have a logarithmic 
branch cut for negative imaginary values of $\bar \omega + \bar k$ and $\bar \omega - \bar k$, respectively. This 
corresponds to the two nonhydrodynamic cuts depicted in Fig.~\ref{fig1}. Moreover, 
the term in the second line of eq.~(\ref{hfunc}) is nonanalytic in the entire strip  $\Im \bar\w < 0$ and $-\bar k < \Re \bar\w < \bar k$ due to the
explicit appearance of $\Im \,\bar \w$. This corresponds to the grey-shaded area of nonanalyticity in Fig.~\ref{fig1}. We note that this nonanalytic
contribution becomes nonperturbatively small for small $\bar\omega$ due to the factor $\sim e^{1/\Im \bar\w}$ in $G(\bar \omega ,\rho=1)$. Therefore, the 
analytic region at $\Im \bar \w \geq 0$ is reached very smoothly, whereas the 
generating function is discontinuous when crossing the $(\Re \bar \w)^2 = {\bar k}^2 $ lines. In contrast to poles and branch-cuts, the analyticity in this strip is also mild
in the sense that a contour integral around a region of area $A$ is proportional to $A$.

The converse of the above statement is not true: the full retarded correlation functions can show additional nonanalytic features that are not visible in the
generating function $H$. There are singular points, arising from the zeroes of the denominators of eqs.~ (\ref{shear}) and (\ref{sound}). These special points
are embedded in the strip of mild nonanalyticity, but they give rise to 
pole-like structures in the sense that they give a finite contribution even when $A$ goes to zero, provided that 
the special point lies within $A$ (see Fig.~\ref{fig1}). Some of these correspond to the hydrodynamical modes
in the model. Indeed, the location of such a special point, in shear channel for example, is given for small $k$ by
\begin{align}
\w_{shear} = - i \frac{\eta}{sT}k^2 + \mathcal{O}(k^4)\, ,
\label{eq55}
\end{align}
as expected from hydrodynamical gradient expansion.  We note that for this result, as for any expression derived in a gradient expansion, 
the nonanalytic parts of the generating function $H$ cannot contribute because of the nonperturbative suppression factor.

\subsection{Ambiguities in the analytical structure}
To obtain correlation functions in the time domain, an inverse Fourier transformation needs to be taken
\begin{align}
G_R^{\alpha \beta,\gamma \delta}(t, k) = \int_{-\infty}^{\infty} \frac{d\omega}{2\pi} e^{- i \w t}G_R^{\alpha \beta,\gamma \delta}(\w, k)\, .
\label{FT}
\end{align}
This expression is typically evaluated by 
completing the contour of the $\omega$-integration along a path at negative complex infinity,  and writing the result as the 
sum of contour integrals around the nonanalytic structures in the negative complex half plane. 
In the present case, however, this standard strategy seems difficult to follow as instead of simple cuts and poles,
the generating function $H$ in eq.(\ref{eq55}) and, a fortiori, the
retarded correlation functions are nonanalytic in an entire two-dimensional region as sketched in Fig.~\ref{fig1}. 

\subsubsection{The analytically continued generating function $H_a$}

A better strategy for calculating (\ref{FT}), that is more practical and more physically revealing 
is to note that the correlation function is analytic in the upper complex half-plane and along the contour
of integration in eq.~(\ref{FT}). Therefore, for the purposes of calculating measurable quantities like (\ref{FT}), 
we may replace the correlation function in the lower complex half-plane with the analytic continuation of the 
function from the upper complex half-plane. The nonanalytic structure of the correlation functions $G^{\alpha \beta,\gamma \delta}(\w, k)$ and
their generating function $H$ in the lower complex half-plane are thus ambiguous to the extent to which the nonanalytic structures 
arising in $H$ can be substituted by an analytic continuation from the upper half-plane.

As the nonanalytic part of the function has already been separated in eq.~(\ref{hfunc}), an
analytic continuation of $H$ from the upper half plane is found simply by removing the nonanalytic part form eq.~(\ref{hfunc}), 
\begin{align}
H_a(\bar \w,\bar k,\rho)=&\frac{-1}{4\bar k} \left[ \frac{\bar \omega-x}{\rho} G(\bar \omega-x,\rho)\right] \bigg\rvert_{-\bar k}^{\bar k}\, .
\label{eq57}
\end{align}
Here, the subscript $a$ stands for analytic continuation. The function $H_a$ contains incomplete gamma functions with 
logarithmic branch cuts whose paths are arbitrary as long as their endpoints
are fixed to $\bar \w= \pm \bar k$ and to negative complex infinity.  Here, we adopt the simplest, but ambiguous choice of continuing 
the complex gamma function to the full complex plane, resulting in branch cuts at $\bar \omega = \pm \bar k + i\bar y$, for real $\bar y \leq 0$.  
So, $H_a$ shows the nonhydrodynamic cuts depicted in Fig.~\ref{fig1}, but unlike $H$, these cuts do not bracket a two-dimensional
strip of mild nonanalyticity.

 \begin{figure}[h]
\includegraphics[width=.4 \textwidth]{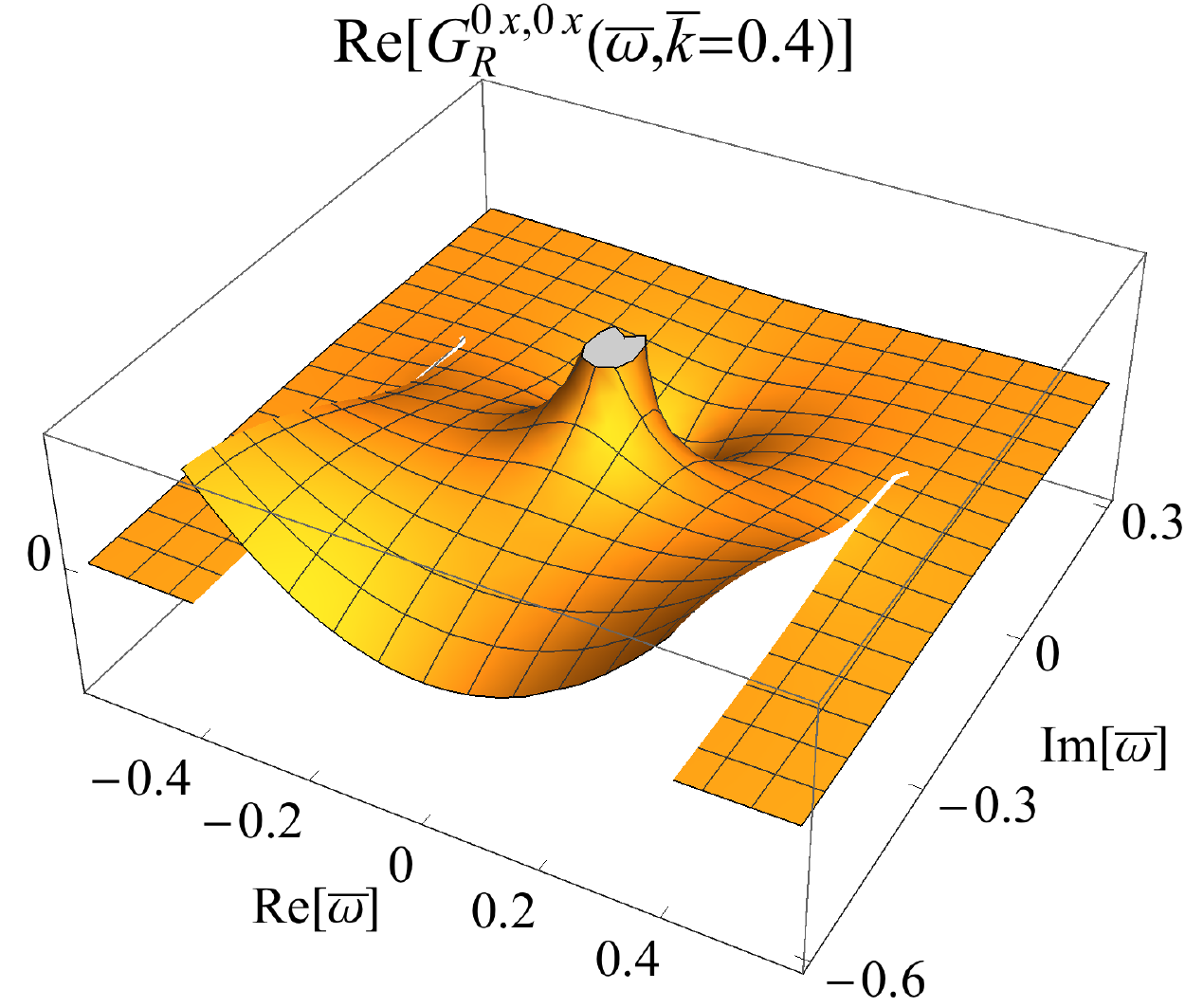}
\includegraphics[width=.4 \textwidth]{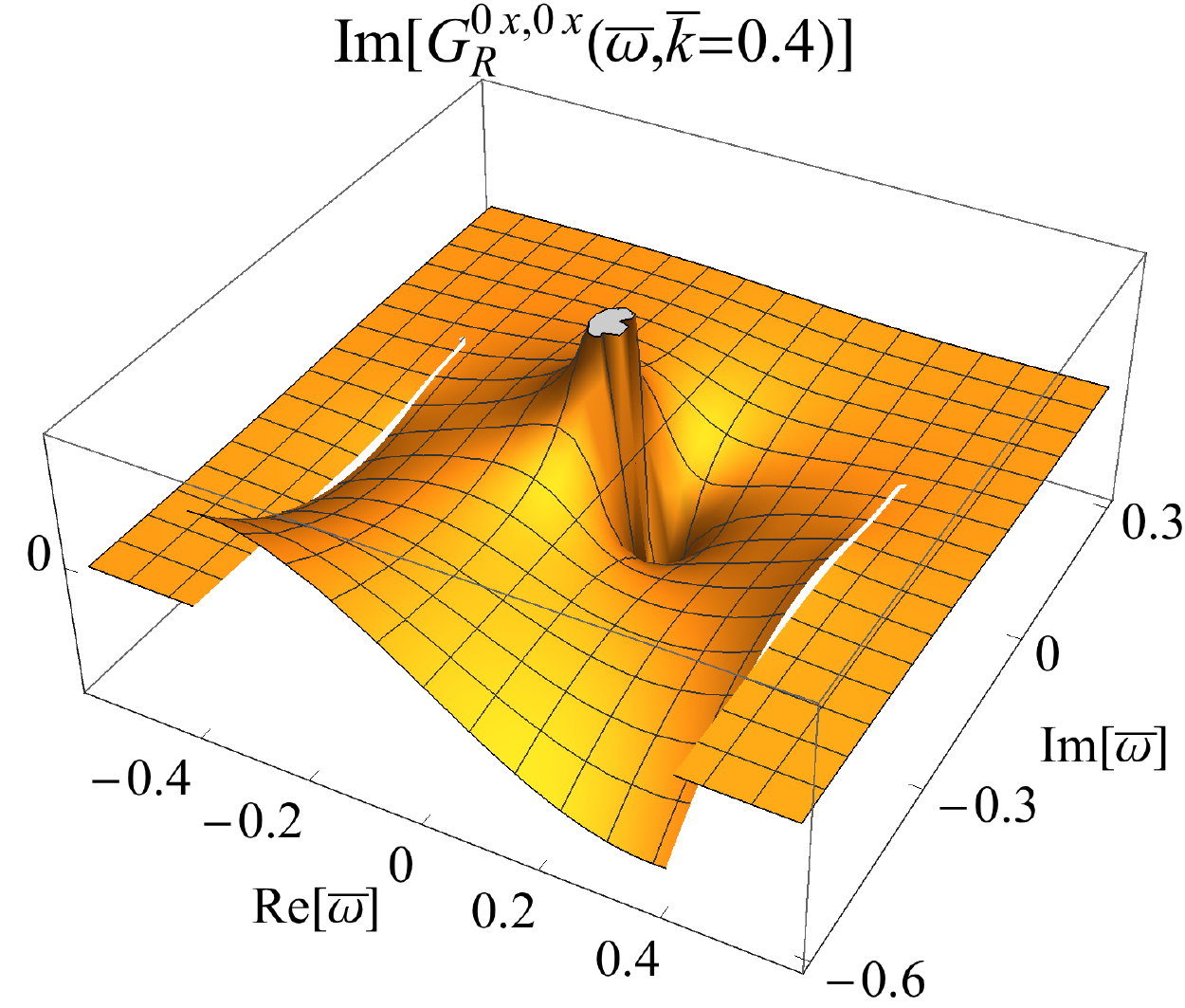}
\caption{The real (left plot) and imaginary (right plot) part of the shear channel retarded correlation function, $G_R^{0x,0x}(\bar{\w},\bar k)$, evaluated for $\bar k = 0.4$
and plotted as a function of complex $\bar\w$. The function $G_R^{0x,0x}(\bar{\w},\bar k)$ is calculated according to eq.~(\ref{shear})  with integral moments evaluated according to
(\ref{eq48}) from the generating function $H_a$ in (\ref{eq57}). 
}
\label{fig4}
\end{figure}

To visualize how the analytic structure of $H_a$ shapes that of retarded correlation functions, we plot in Fig.~\ref{fig4} the real and imaginary part of the shear channel $G^{0x,0x}_R$,
calculated from $H_a$.
This correlation function clearly shares with $H_a$ the two branch cuts that run in the negative imaginary 
half plane along $\Re(\bar\w)=\pm \bar k$ from zero to
complex negative infinity. Closer inspection also reveals that the discontinuity across these branch cuts is exponentially small for small $\Im(\bar\w)$, as expected from
the factor $\exp\left[i\rho/\bar\w \right]$ in (\ref{G}). In addition, there is a prominently visible structure of neigboring peak and trough close to $\Re(\bar\w)=0$ at negative $\Im(\bar\w)$,
whose orientation is rotated by $\pi/2$ between the real and imaginary part of $G^{0x,0x}_R$. This is the tell-tale signature of a simple pole $\propto 1/(\bar\w + i\, {\rm const})$
in the complex plane. The precise location of this hydrodynamic pole will be discussed in the following. In the gradient expansion, it is given of course by (\ref{eq55}).

 \subsubsection{Deforming the branch cuts}
 The purpose of this section is to show that in general, the presence or absence of hydrodynamic poles  
 in the lower imaginary half plane of $G^{\alpha \beta,\gamma \delta}(\w, k)$ is not indicative of the onset or disappearance of fluid dynamic behavior. 
 
 To set the stage of this discussion, we note first that the same physical response $G_R^{\alpha \beta,\gamma \delta}(t, k)$ in the time
 domain can be encoded in different analytical structures  $G_R^{\alpha \beta,\gamma \delta}(\w, k)$ in the complex frequency domain. 
This was illustrated already by showing that constructing $G^{\alpha \beta,\gamma \delta}(\w, k)$ from 
the generating function $H$ in eq.~(\ref{hfunc}) or from $H_a$ in eq.~(\ref{eq57}) yields physically identical responses $ G_R^{\alpha \beta,\gamma \delta}(t, k)$
while the analytic structure of $G^{\alpha \beta,\gamma \delta}(\w, k)$ is qualitatively different for both cases in the sense that it has a two-dimensional 
region of mild nonanalyticity if constructed from $H$, but not if constructed from $H_a$. In the present section, we consider formulations of the latter
kind, for which $G_R^{\alpha \beta,\gamma \delta}(\w, k)$ is given in terms of branch cuts and poles only. In particular, the construction of $ G_R^{\alpha \beta,\gamma \delta}(t, k)$
from the generating function $H_a$ is technically advantageous, since the contour of the integration (\ref{FT}) can
be closed by encircling the branch cuts going from $\pm k$ to $\pm k - i \infty$ 
and encircling any hydrodynamical poles $\w_i$ that may be found in the given channel, 
\begin{align}
G_R^{\alpha \beta,\gamma \delta}(t, k)&  = - 2\pi i  \sum_i {\rm Res}(\w_i)e^{- i \w_i t} \label{FT2}\\
&+ 2 \Im e^{- ikt}\int_{-\infty}^{0}d y e^{y t} {\rm Disc} G^{\alpha \beta,\gamma \delta}(k + i y, k) \nonumber,
\end{align}
As we shall illustrate in the following with an explicit
construction, only the sum of the pole and cut contributions on the right hand side of (\ref{FT2}) is physical. The relative weight of both terms depends on the
orientation of the branch cuts in the lower complex half plane, which is a purely technical choice without unambiguous physical interpretation. 

\begin{figure}[h]
\includegraphics[width=.4 \textwidth]{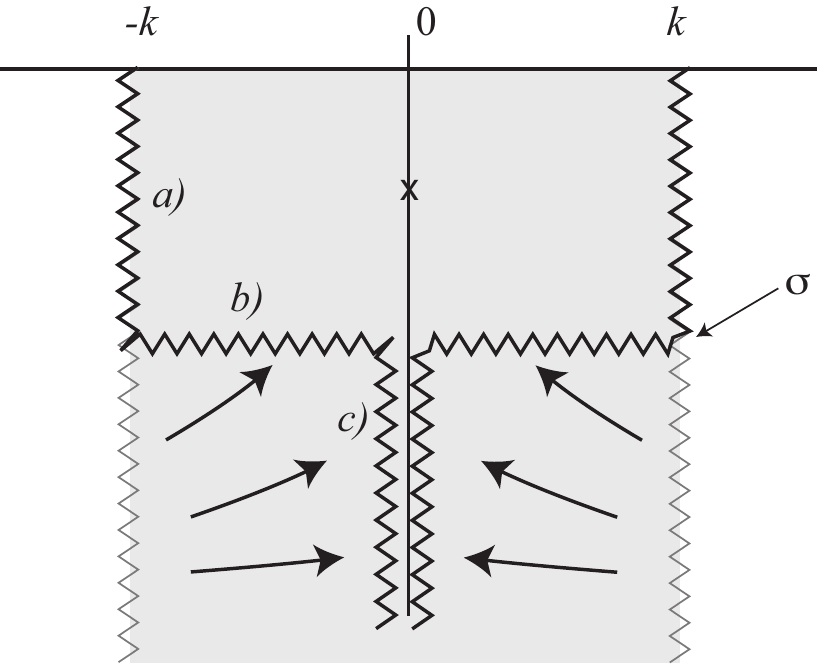}
\caption{A particular deformation of the branch cuts of the generating function $H_a$, defined in eq.~(\ref{eq59}) and the surrounding text. 
}
\label{fig4b}
\end{figure}

\begin{figure*}
\includegraphics[width=0.4\textwidth]{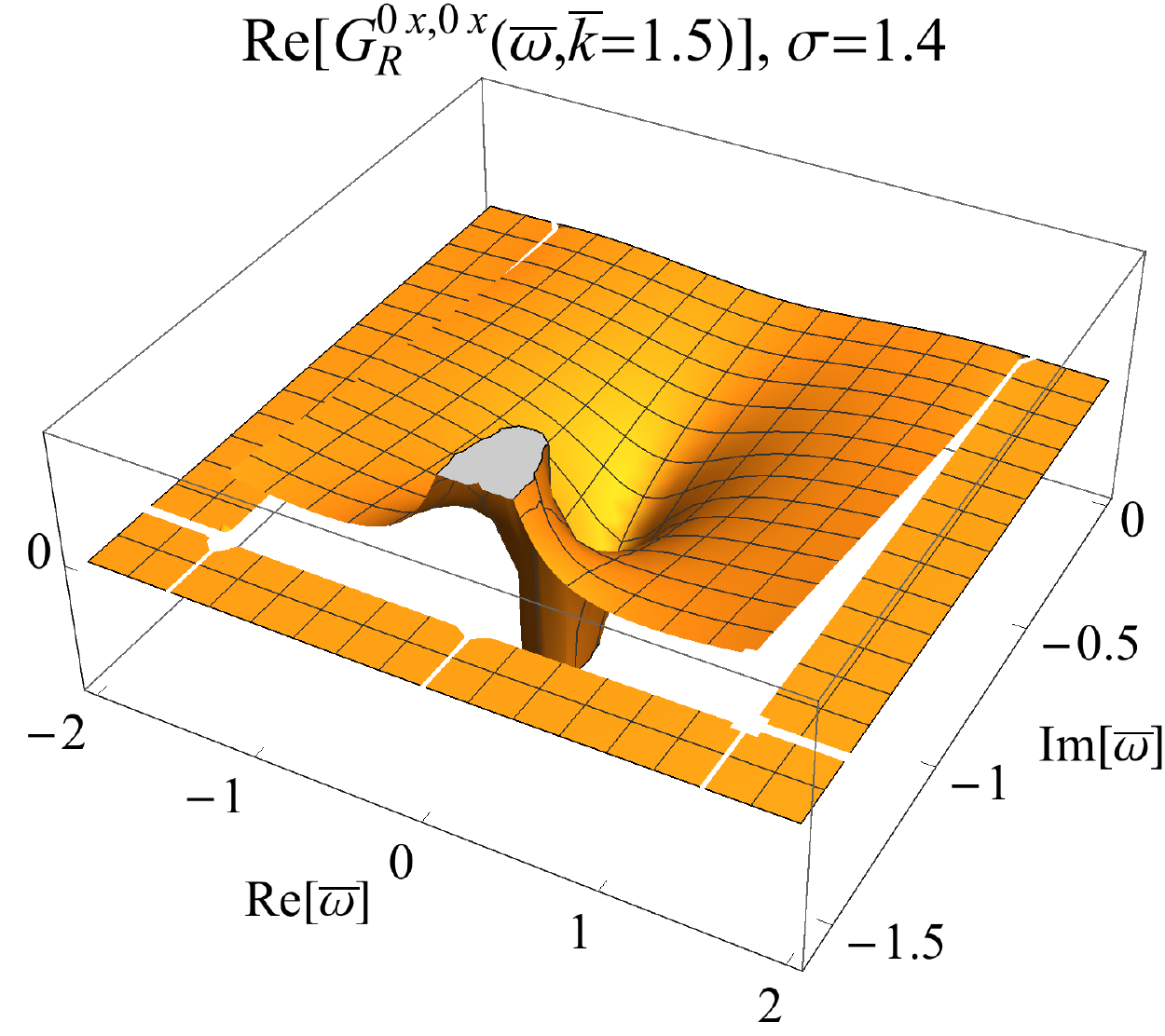}
\includegraphics[width=0.4\textwidth]{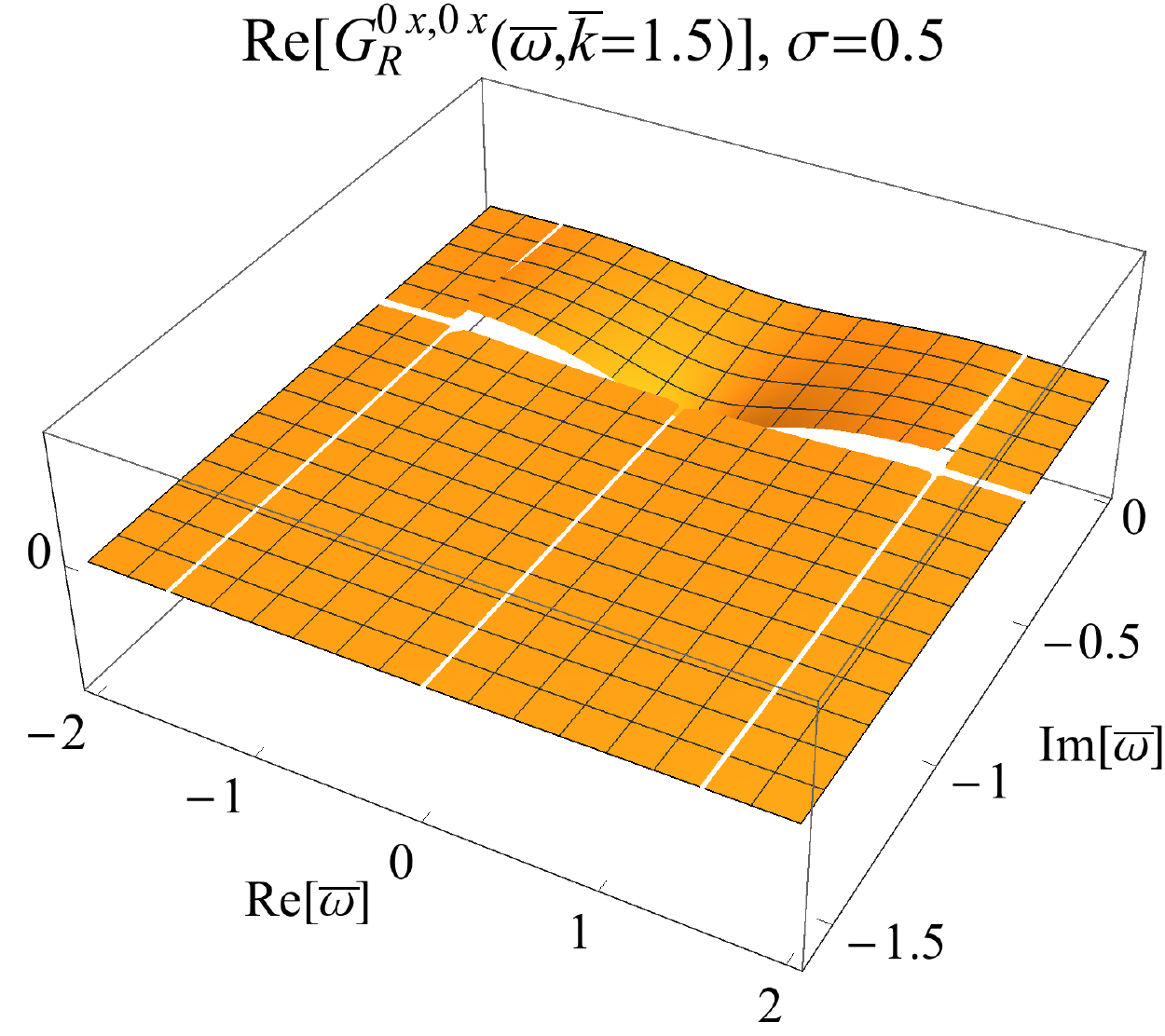}
\includegraphics[width=0.4\textwidth]{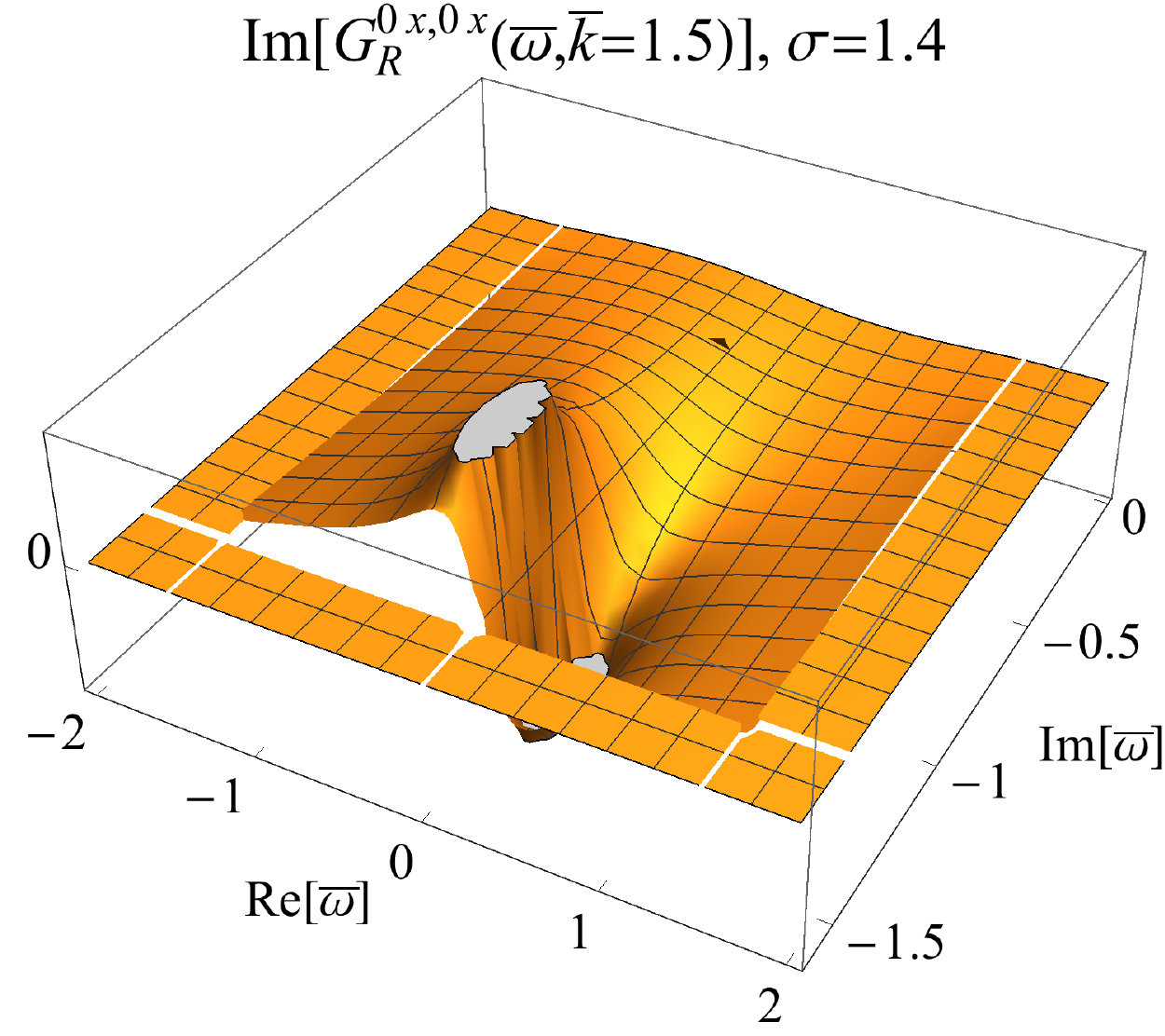}
\includegraphics[width=0.4\textwidth]{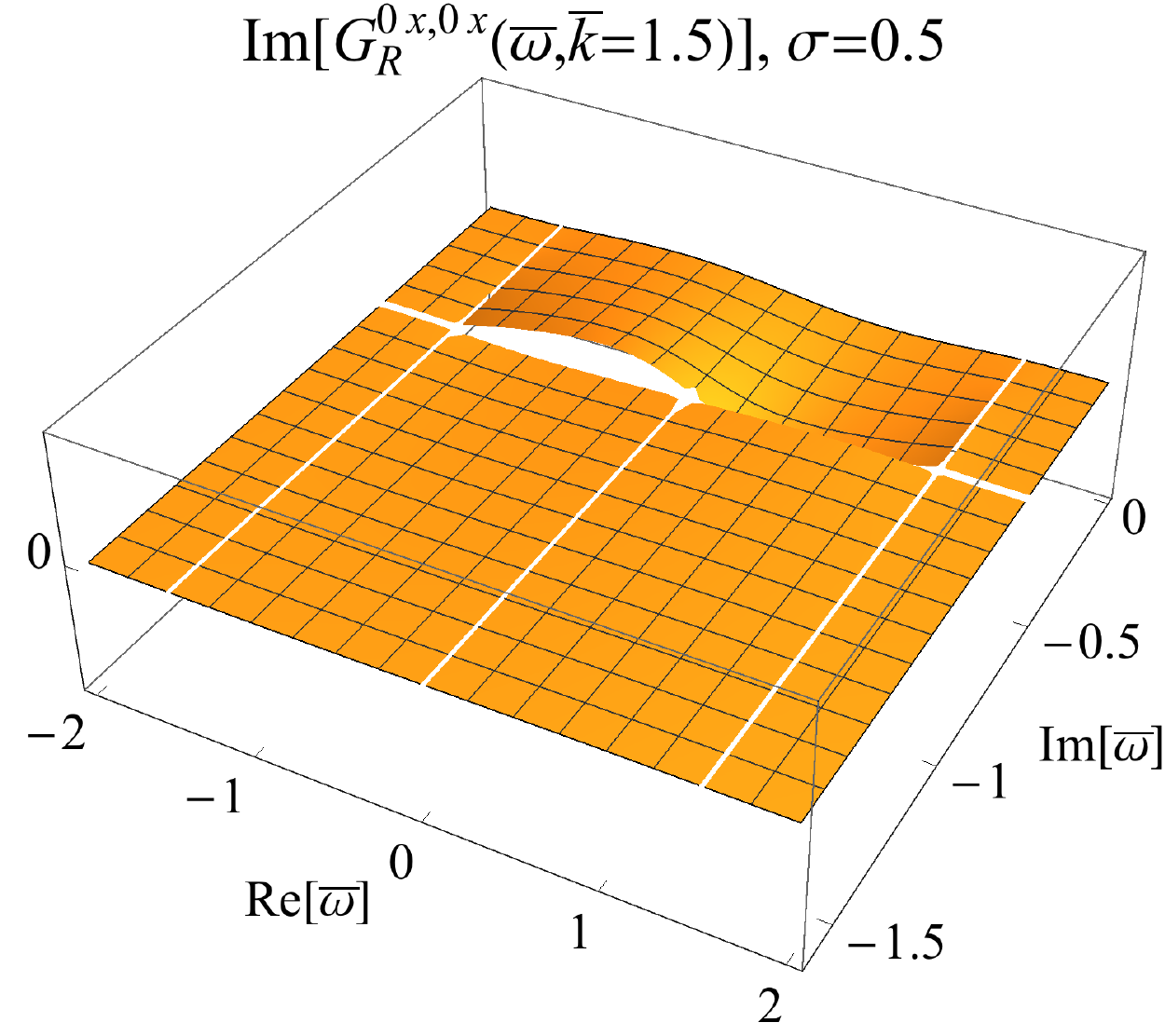}
\caption{
The real and imaginary part of the retarded correlation function $G_R^{0x,0x}(\bar \w,\bar k = 1.5)$, evaluated for choices of the logarithmic branch cuts depicted in Fig.~\ref{fig4b}.
Left hand side: If the two branch cuts are joined at imaginary
depth $\sigma = 1.4$, then the shear pole is clearly visible above the cut. Right hand side: for a different choice of $\sigma = 0.5$, the pole moves under the cut while the physical properties in 
the time domain remain by construction identical to those defined by the correlation functions on the left hand side.  
}
\label{fig5}
\end{figure*}

If calculated from $H_a$, all integral moments entering the correlators $G_R^{\alpha \beta,\gamma \delta}(\w, k)$ can be expressed in terms 
of rational functions and rational functions times $G_+ = \frac{e^{\frac{i}{\bar \w + \bar k}}}{\bar \w + \bar k}\Gamma\left(0, \frac{i }{\bar \w + \bar k}\right)$
or $G_- = \frac{e^{\frac{i}{\bar \w - \bar k}}}{\bar \w - \bar k}\Gamma\left(0, \frac{i }{\bar \w - \bar k}\right)$. The branch cuts of $G_R^{\alpha \beta,\gamma \delta}(\w, k)$
are therefore determined by the logarithmic branch cuts of $\Gamma\left(0, \frac{i }{\bar \w \pm \bar k}\right)$. 
To be specific, we consider now a 
particular deformation of these branch cuts, sketched in Fig.~\ref{fig4b} and defined by the replacement
\begin{align}
	&\Gamma\left(0, \frac{i }{\bar \w + \bar k}\right) = R_{\rm reg} (\bar \w + \bar k) - \log \left(\frac{i }{\bar \w + \bar k}\right) \nonumber \\
	&\longrightarrow \Gamma\left(0, \frac{i }{\bar \w + \bar k}\right) + \log \left(\frac{i }{\bar \w + \bar k +i\sigma}\right) \nonumber \\
	& \qquad  - \log \left(\frac{-1 }{\bar \w + \bar k +i\sigma}\right) +  \log \left(\frac{-1 }{\bar \w + i\sigma}\right)  \nonumber \\
	 & \qquad - \log \left(\frac{i }{\bar \w + \bar k +i\sigma}\right)\, .
	 \label{eq59}
\end{align}
Here, $R_{\rm reg}$ denotes the regular part of the $\Gamma$-function. In the replacement (\ref{eq59}), the logarithm in the second line of (\ref{eq59}) cancels part of the branch cut of 
the $\Gamma$-function, such that only the segment a) in Fig.~\ref{fig4b} remains. The two logarithms in the third line of (\ref{eq59}) combine to the segment b) in Fig.~\ref{fig4b},
and the logarithm in the last line corresponds to segment c). 
We deform the branch cut of $G_-$ symmetrically (see Fig.~\ref{fig4b}), so that both branch cuts meet at $\bar\w =  - i \sigma$ on the imaginary axis, and are then continued 
on top of each other up to 
complex imaginary infinity. This deformation leaves the generating function unchanged for $\Re(\bar \w) \geq 0$ and it therefore encodes the same physics.

  In Fig.~\ref{fig5}, we plot the real and imaginary parts of the retarded correlation function in the shear channel for this choice of branch cuts.\footnote{We note that our
  construction of these branch cuts in (\ref{eq59}) involves pairs of logarithmic cuts that cancel each other outside a finite segment. For instance, the two terms in the 
  third line of (\ref{eq59}) extend both to $\bar \w = -i\sigma + \infty$ but they cancel each other for $\Re(\bar \w) > \bar k$. The numerical evaluation shown in Fig.~\ref{fig5}
  does not attribute values to these lines along which logarithm contributions cancel each other, even though the correlation function is regular there.}  Depending on the
  depth $-i\sigma$ in the complex $\bar \w$-plane at which the two branch cuts are joined, the shear pole is either clearly visible (left hand side of Fig.~\ref{fig5}), or it 
  disappears under the branch cut. We emphasize that while both choices of $\sigma$ lead to qualitatively different features in the analytical structure of 
  $G_R^{\alpha \beta,\gamma \delta}(\w, k)$, they are physically equivalent in the sense that they give rise to identical physical responses $G_R^{\alpha \beta,\gamma \delta}(t, k)$
  in the time domain. In this sense, the appearance or disappearance of a hydrodynamic-like pole is related to purely technical and physically ambiguous choice of branch cut 
  and it therefore cannot be related to the onset of fluid dynamic behavior.

  \subsubsection{Differences between the cases $\xi = 0$ and $\xi > 0$}
As explained in Appendix~\ref{appb}, eq.~(\ref{eqb3}), the integral moments (\ref{eq30}) that define retarded correlation functions for the case
of a scale-independent relaxation time, $\xi=0$, can be written in terms of rational functions 
of $\bar\w$ and $\bar k$, and in terms of rational functions times the difference of logarithms 
\begin{align}
 &\propto \left[   \log\left( \w- k+i/\td \right) - \log\left( \w+ k+i/\td \right)  \right]\, .
 \label{eq58}\\
 &\qquad \qquad \hbox{[for the case $\xi = 0$]} \nonumber
\end{align}
This is consistent with the qualitative argument leading to (\ref{eq14}). As a consequence, for $\xi = 0$, the
retarded correlation functions share the nonanalytic structure of (\ref{eq58}). 

According to the standard definition, the branch cuts of the logarithms in (\ref{eq58}) start at $\w = - i/\td \pm k$ and they run parallel to
the real axis to $\w = -i/\td - \infty$. Therefore, they cancel each other outside the range $-k \leq \Re\w \leq k$,
and this gives rise to the nonanalytic segment sketched in Fig.~\ref{fig3}. 
However, the two logarithmic branch cuts of (\ref{eq58}) could also be deformed to run parallel to the imaginary axis from $\w =  \pm k -i/\td$ to negative complex infinity, $\w = \pm k -i\infty$.

These two ways of orienting the branch cuts of (\ref{eq58}) are reminiscent of the two choices of branch cuts for $H_a$ depicted in Fig.~\ref{fig5} and discussed for $\xi =1$ in the 
previous subsections. However, there are marked physical differences between the cases $\xi =0$ and $\xi > 0$:

First, for $\xi = 0$, the branch cuts can be oriented such that for sufficiently small $k$, hydrodynamic poles are the unique nonanalytic structure closest to the real axis,
 thus determining the late-time behavior of retarded correlation functions, see eq.(\ref{FT2}). In contrast, for $\xi = 1$, the branch cuts start always at $\bar \w = \pm \bar k$, 
and for a gradient expansion around $\bar k=0$, poles and the starting point of branch cuts are not separated.
This observation is related to the finding that the gradient expansion for the position of the pole converges for the case $\xi = 0$ (for instance, 
$\w_{\rm shear}(k)\vert_{\xi = 0} = \textstyle\frac{-i}{\td}  + \textstyle\frac{ik}{\tan(\bar k)}$~\cite{Romatschke:2015gic}), while it is an asymptotic series for $\xi = 1$ (see discussion of Fig.~\ref{fig4?} below).

Second, for $\xi = 0$, the branch cuts in (\ref{eq58}) can cancel each other outside a finite segment.  As illustrated in Fig.~\ref{fig4?}, this is not possible for the case $\xi = 1$. 
If one deforms the branch cuts of $H_a$ so that they lie on top of each other from $\bar \w = -i\sigma$ up to $\bar \w = -\infty$, they will not cancel exactly. Rather,
along the line of overlapping branch cuts, there will be a discontinuity
\begin{align}
	&H_a^{\rm Right} (i\Im(\bar \w)+\epsilon, \bar k) - H_a^{\rm Left} (i\Im(\bar \w)-\epsilon, \bar k) \nonumber \\
	& = \frac{i\pi}{\bar k} \left(\exp\left[\frac{i}{-\bar k + \bar \w}\right] - \exp\left[\frac{i}{\bar k + \bar \w}\right] \right)\, ,
\end{align}
where $H_a^{\rm Right}$, $H_a^{\rm Left} $ denote analytically continued branches of $H_a$ as defined in the caption of Fig.~\ref{fig4?}. 
 \begin{figure}[h]
\includegraphics[width=.4 \textwidth]{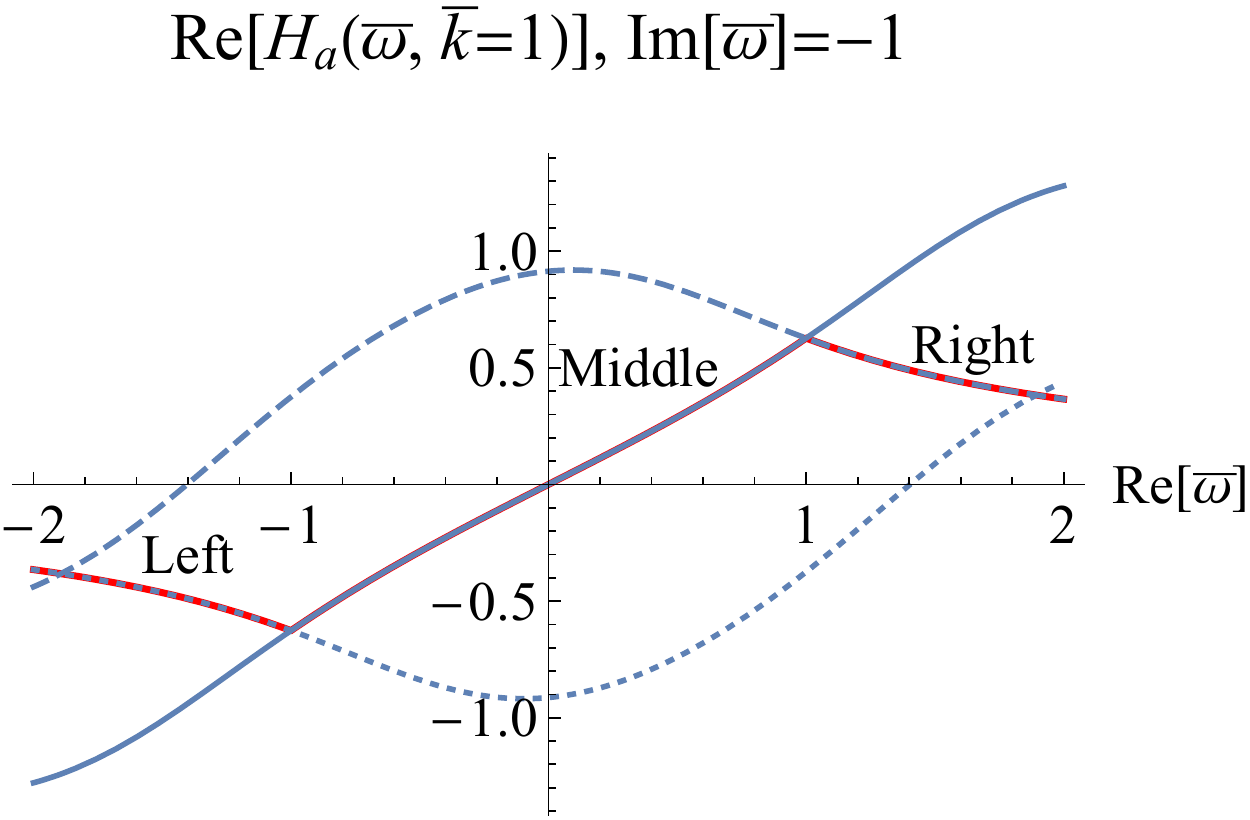}
 \vskip .6cm
\includegraphics[width=.4 \textwidth]{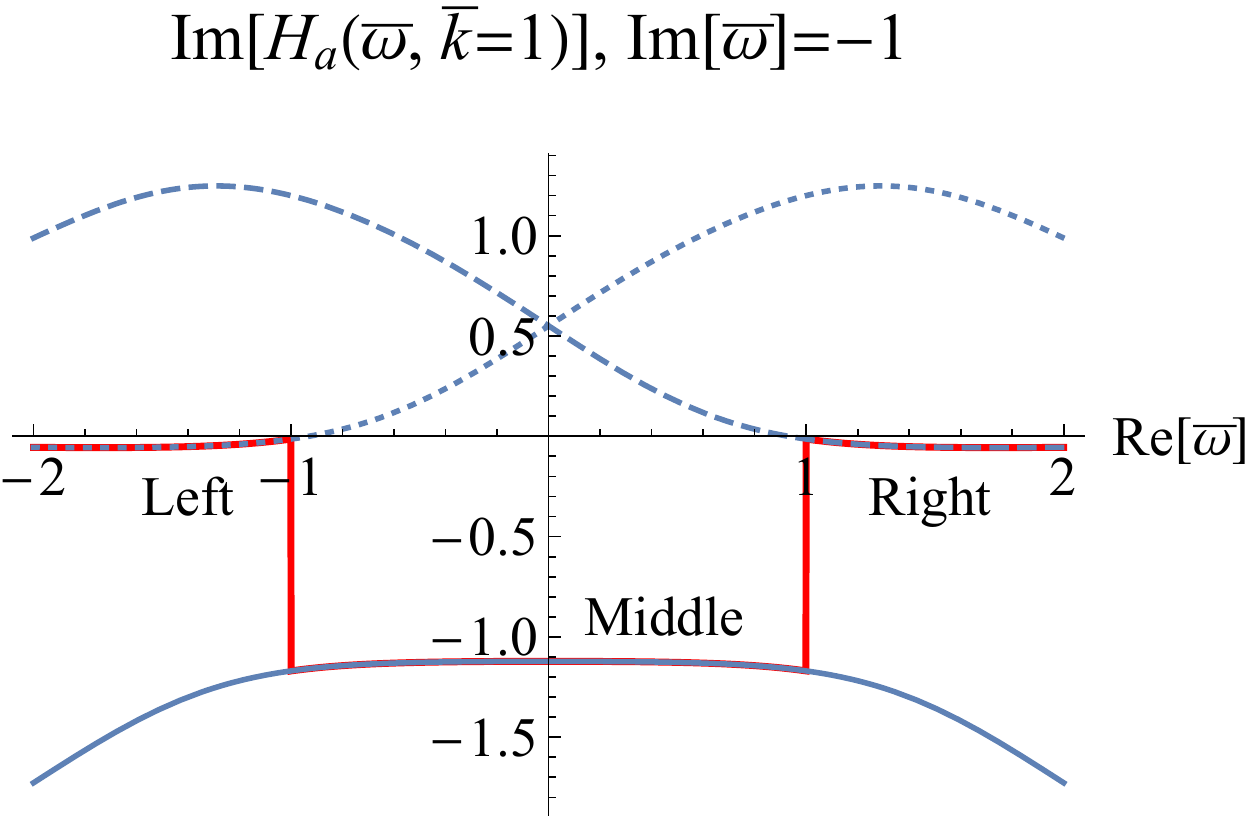}
\caption{Real and imaginary part of the function $H_a$, defined in the Left  ($\Re(\bar \w) < - \bar k$), Middle  ($ - \bar k < \Re(\bar \w) < \bar k$) and Right 
($ \bar k < \Re(\bar \w)$) part of the complex plane and analytically continued into the other regions of the $\w$-plane. The red curves indicate the function
for a choice of the two branch cuts that pass $\Re(\bar \w) = \pm 1$ at $\Im(\bar\w) = -1$. Deforming the branch cuts amounts to varying the positions
$\Re(\bar\w)$ at which the different analytical patches of $H_a$ are interfaced. 
}
\label{fig4?}
\end{figure}

\section{Retarded correlation functions $G_R^{\alpha \beta,\gamma \delta}(t, k)$ in the time domain}
\label{sec5}
In this section, we utilize our understanding of the nonanalytic structures of $G_R^{\alpha \beta,\gamma \delta}(\w, k)$ in the frequency domain for a discussion 
of the physical response  $G_R^{\alpha \beta,\gamma \delta}(t, k)$ in the time domain. The connection between both is given by eq.(\ref{FT2}). 

In general, with small but increasing $k$, the pole contributions to $G_R^{\alpha \beta,\gamma \delta}(t, k)$ in (\ref{FT2}) move deeper into the complex plane and they start 
being cancelled more efficiently by the discontinuities from the branch cuts. While only the sum of these nonanalytic contributions has unambiguous physical meaning,
the separate determination of both, the poles and their residues,  and the discontinuities along the branch cuts is needed in practice for a discussion of the full physical 
response in the time domain. In the following, we discuss these nonanalytic contributions separately for the specific choice of the generating function $H_a$
in (\ref{eq57}) with branch cuts taken along $\bar\w = \pm \bar k + i\, y\, \td$, $y \in [0, -\infty ]$.

\subsection{The location of the hydrodynamic poles in the shear and sound channel}

 \begin{figure}[t]
\includegraphics[width=.48 \textwidth]{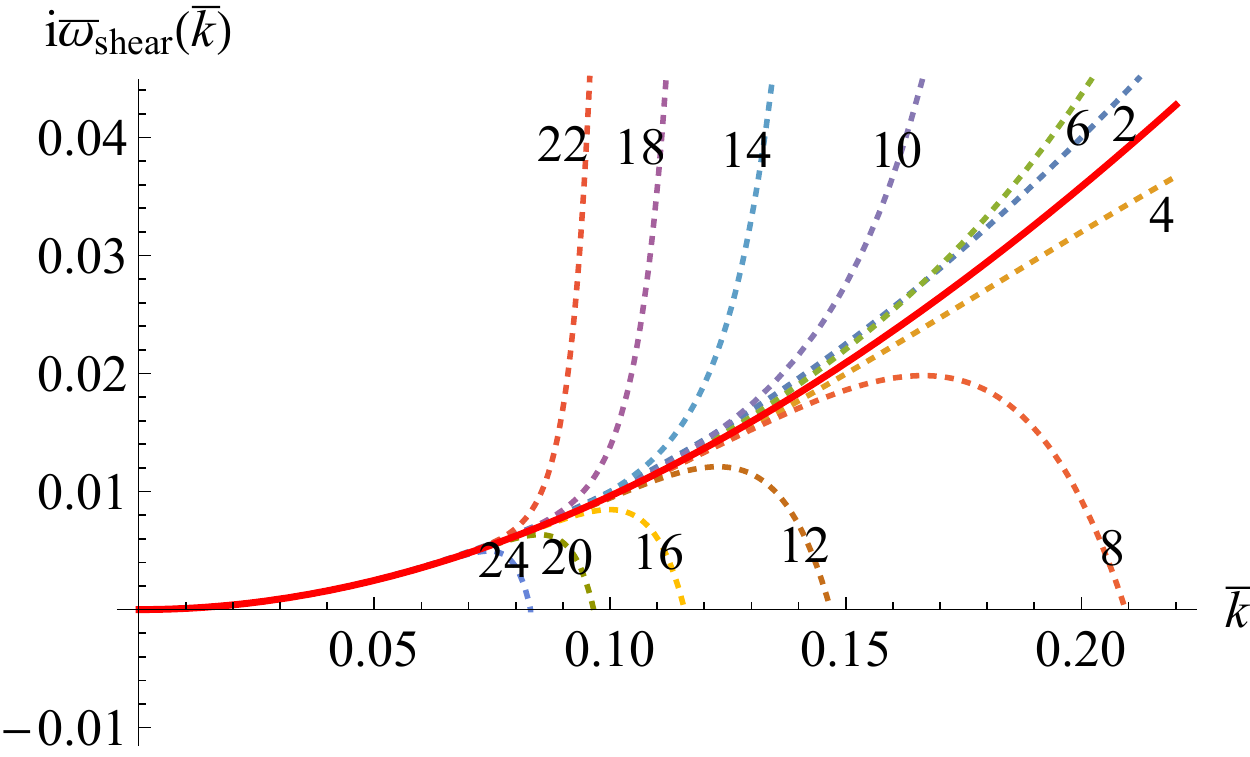}
\caption{The pole $\bar{\w}_{\rm shear}(\bar{k})$ of the shear channel correlator $G_R^{0x,0x}(\bar{\w},\bar k)$ (red curve) compared to gradient expansions 
(\ref{eq60}) of $\bar{\w}_{\rm shear}(\bar{k})$ up to power $\bar{k}^{2N}$. The integers $2N$ on the dashed lines denote the highest power $\propto k^{2N}$ included
in the gradient expansion. 
}
\label{fig6}
\end{figure}

\subsubsection{The pole in the shear channel}
 
 The pole $\bar{\w}_{\rm shear}(\bar{k})$ of the retarded correlation function $G^{0x,0x}(\bar{\w},\bar k)$ is defined implicitly in terms of the zero 
 of the nontrivial denominator in eq.~(\ref{shear}),
 \begin{align}
 	2 - 3\,  I^{2\xi,2,0}\left(-i\, \bar{\w}_{\rm shear}(\bar{k}), \bar{k}\right) \equiv  0\, .
 \end{align}
 This equation can be solved numerically without any recourse to the gradient expansion. Alternatively, it can be solved by determining the first $N$ coefficients $b_i$ in a gradient expansion 
 \begin{align}
 i\, \bar{\w}_{\rm shear}(\bar{k}) = \sum_{j=1}^N b_j\, ({\bar k})^{2j}\, .
 \label{eq60}
 \end{align}
In Fig.~\ref{fig6}, the exact solution is compared with this gradient expansion. With increasing orders $\propto {\bar k}^{2N}$, the gradient expansion is seen to deviate from the exact 
result at smaller and smaller $\bar k$. This illustrates that the gradient expansion is an asymptotic expansion that does not possess a finite radius of convergence. 

For large ${\bar k}$, the hydrodynamic pole moves deep into the complex plane
\begin{align}
\bar \w_{shear} &\approx -i  \sqrt{\frac{2}{\pi }}\bar k^{3/2} + \frac{i \bar k}{2 \pi }+\mathcal{O}(\bar k^{1/2})\, .
\end{align}

\begin{figure}[h]
\includegraphics[width=.48 \textwidth]{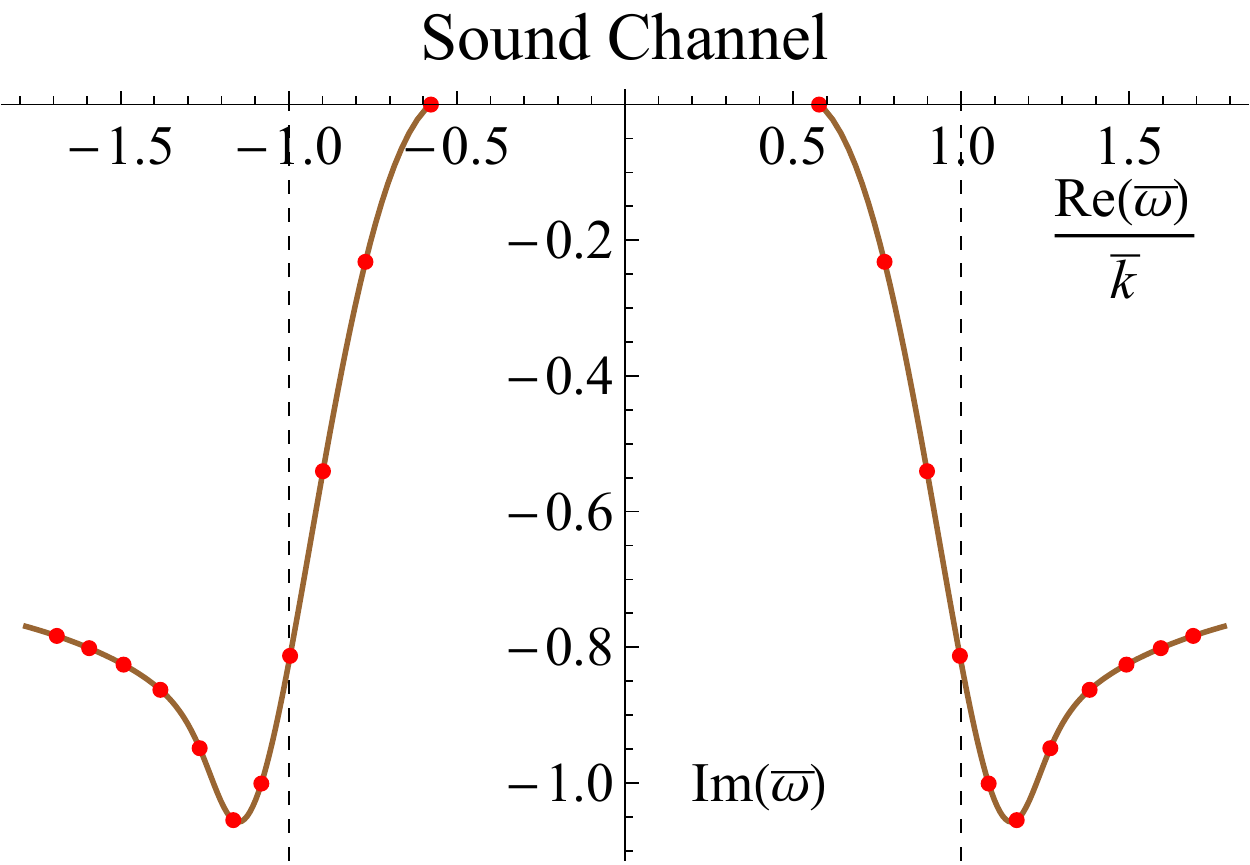}
\caption{The location of the two sound poles in the lower complex plane. 
The red dots are at values of $\bar k = 0,1,2,\ldots$. For 
large wavelengths, the location and residue are described by hydrodynamic 
gradient expansion. For large $\bar k$ the poles move to large real frequencies.}
\label{fig8}
\end{figure}

\subsubsection{The sound channel}
In close analogy to the discussion of the pole in the shear channel, the poles in the sound channel can be determined in terms of the zeros of the 
denominator of (\ref{sound}). While the pole in the shear channel is purely imaginary, the pair of sound poles start at finite real values $\bar \w_{\rm sound}(\bar k=0) = \pm c_s = \pm \textstyle\frac{1}{\sqrt{3}}$ before diving into the negative  
imaginary half plane. The full numerical solution is shown in Fig.~\ref{fig8}.

We note that branch cuts can be chosen such that hydrodynamic poles disappear below the cut in one channel while they do not disappear
in another channel. Here, this is the case for the choice of branch cuts in $H_a$ along the imaginary axis. For this choice, the shear pole will remain
visible for all $\bar k$, while the sound pole disappears at $\bar k = 4$, see Fig.~\ref{fig8}. This is yet another illustration of the general statement that
there is no unambiguous relation between the existence of hydrodynamic poles in the retarded correlator and the persistence of fluid dynamic  
behavior.

We further observe with curiosity that the positions of the sound poles move first away from the real axis, before they move closer to the
real axis again, see Fig.~\ref{fig8}. We note that other cases are known in the literature where a pole moves closer to the real axis
with increasing $k$, see e.g. Ref.~\cite{Chesler:2011nc}.
The asymptotic large-$k$ behavior is given by
\begin{align}
\bar \w_{sound} &\approx  \frac{1}{\sqrt{ \pi }} \bar k^{3/2} - \frac{\sqrt \pi}{6}\bar k^{1/2} - \frac{2 i}{3}+\mathcal{O}(\bar k^{-1/2}),
\end{align}
%

\subsection{Contributions of the branch cuts to $G_R^{\alpha \beta,\gamma \delta}(t, k)$}
\begin{figure*}
\includegraphics[width=.7 \textwidth]{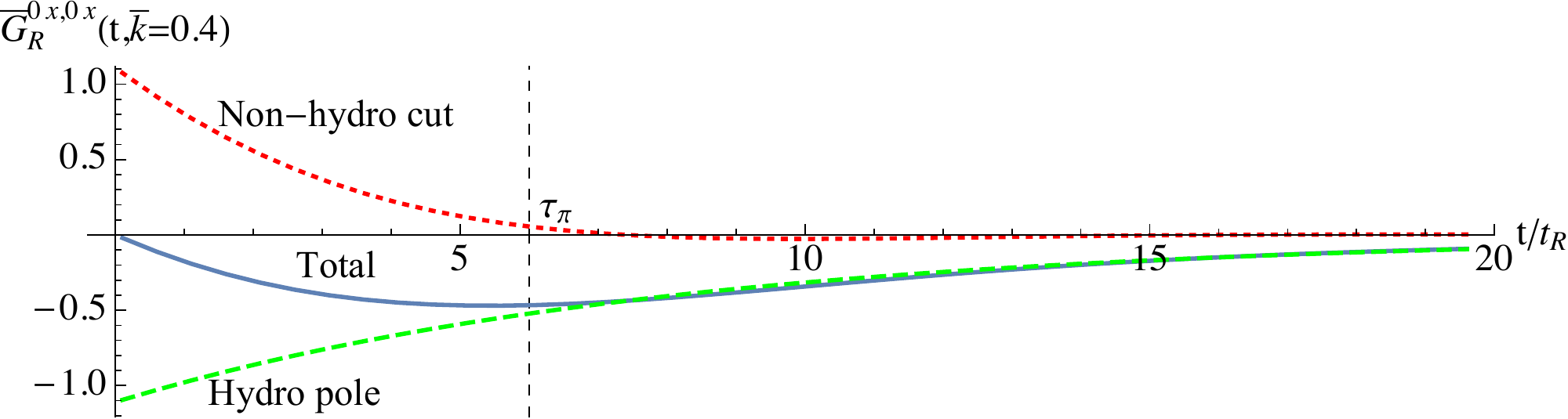}
\includegraphics[width=.7 \textwidth]{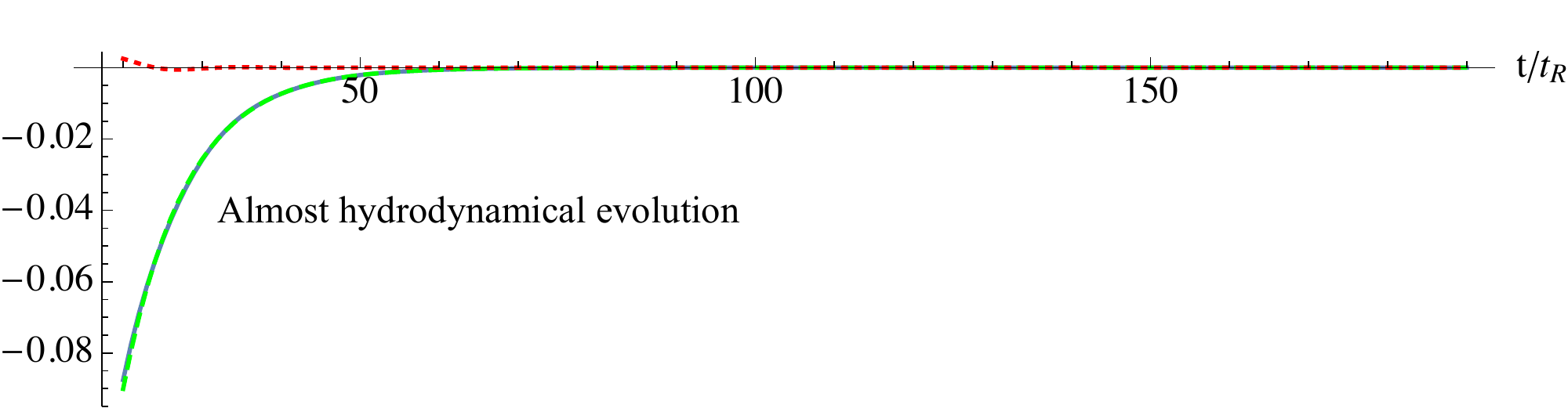}
\includegraphics[width=.7 \textwidth]{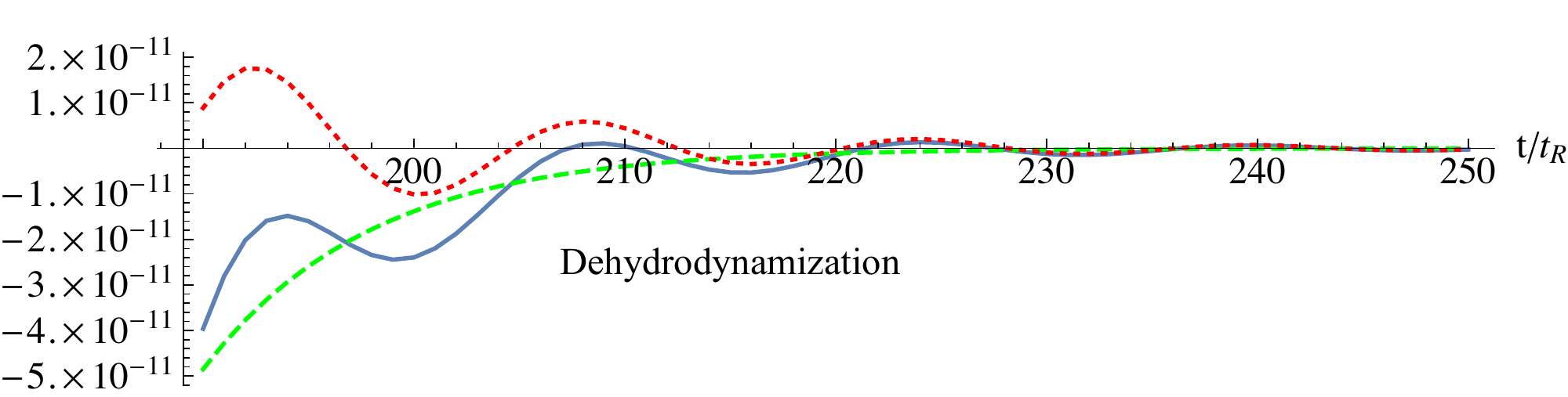}
\caption{The full physical response $G_R^{0x,0x}(t, k)$ as a function of time for $\bar k = 0.4$ (blue line) and the individual pole (green dashed line) and
cut (red dotted line) contributions that determine $G_R^{0x,0x}(t, k)$ according to eq.~(\ref{FT2}). The dashed vertical line indicates the time-scale 
$\tau_\pi = 6\, \td$ where hydrodynamization is estimated to be accomplished, see eq.(\ref{eq44}).}
\label{fig9}
\end{figure*}

We now combine the information gathered about the nonanalytic structure of $G_R^{\alpha \beta,\gamma \delta}(\w, k)$
to arrive via eq.~(\ref{FT2}) at a qualitative understanding of the time-dependence of the physical
response $G_R^{\alpha \beta,\gamma \delta}(t, k)$. For the shear channel, this time dependence is illustrated with the
numerical results in Fig.~\ref{fig9} that display the three characteristic stages of hydrodynamization, hydrodynamic evolution 
and dehydrodynamization. The following discussion aims at providing an analytic understanding for how these features arise.

For notational simplicity, we work in the following with
\begin{align}
	{\bar G}_R^{\alpha \beta,\gamma \delta} \equiv  \frac{1}{sT} G_R^{\alpha \beta,\gamma \delta}\, .
\end{align}

\subsubsection{The limit $t\to 0$ of the retarded correlation functions}
In the kinetic theories studied here, the physical response at time $t=0$ starts always from
\begin{align}
{\bar G}_R^{\alpha \beta,\gamma \delta}(t=0, k) = 0\, .
\label{eq66}
\end{align}
This can be seen by expanding the exponent in the Fourier transform 
for small $t$,
\begin{align}
{\bar G}_R^{\alpha \beta,\gamma \delta}(t, k) = \int_{-\infty}^{\infty} \frac{d\omega}{2\pi}\left( 1- i \w t + \ldots \right)\, {\bar G}_R^{\alpha \beta,\gamma \delta}(\w, k)\, .
\label{eq67}
\end{align}
One checks explicitly for each channel of interest that $G_R^{\alpha \beta,\gamma \delta}(\w, k)$ falls off like $\propto 1/\w^2$ or faster for large $\w$.
Therefore, the first term in the expansion (\ref{eq67}) can be obtained by closing the integration contour in the positive imaginary half plane where integrals
along closed contours vanish due to the analyticity of $G_R^{\alpha \beta,\gamma \delta}(\w, k)$. This implies that the small-$t$ expansion of
$G_R^{\alpha \beta,\gamma \delta}(t, k) $ starts with a positive power of $t$ and that eq.~(\ref{eq66}) is satisfied. 

According to eq.~(\ref{FT2}), the pole contribution to $G_R^{\alpha \beta,\gamma \delta}(t, k)$ at time $t=0$ is a sum of residues which is nonzero. To 
satisfy (\ref{eq66}), this pole contribution must therefore cancel exactly the contribution from the cut at time $t=0$.
To see this cancellation explicitly at work, we consider the shear channel that has one single pole, and we focus for
simplicity on large $k$. In this limit, the residue of the shear pole of $\bar G^{0x,0x}_R$ is 
\begin{align}
{\rm Res}(\bar \w_{shear}) &\approx \frac{-9i \bar k^3}{4\pi} + \mathcal{O}(\bar k^{3/2})\, .
\label{eq70}
\end{align}
Therefore, the pole contribution to the retarded correlation function (\ref{FT2}) diverges for large $k$ and small $t$.
In the same limit, the cut contribution is sharply peaked around the location of the pole
\begin{align}
{\rm Disc}{\bar G}_R^{0x,0x}(\bar k + i \bar y, \bar k) \approx & \frac{9 i \bar k^4}{4 \pi  (\bar y + i \bar \w_{shear}) ^2+4 \pi  \bar k^2} \nonumber \\
& + \mathcal{O}(\bar y + i \bar \w_{shear})^2\, ,
\end{align}
such that the contribution from the discontinuity for large $k$ and small $t$ reads
\begin{align}
2 & \Im \int_{-\infty}^{0}d y {\rm Disc}{\bar G}_R^{0x,0x}(\bar k + i \bar y, \bar k) = \frac{9 \bar k^3}{2}\, .
\label{eq72}
\end{align}
So, indeed, cut and pole contribution cancel exactly for $t = 0$. Since both are continuous in $t$, they weill cancel partially for short times $t> 0$.

\subsubsection{Hydrodynamization}
In applications of hydrodynamics, it is often assumed that hydrodynamic behavior dominates the evolution of near-equilibrium perturbations on time scales
$t > \tau_\pi$. In the kinetic model studied here, this hydrodynamic shear relaxation time (\ref{eq44}) is $ \tau_\pi = 6 \td$.

\begin{figure}
\includegraphics[width=0.4\textwidth]{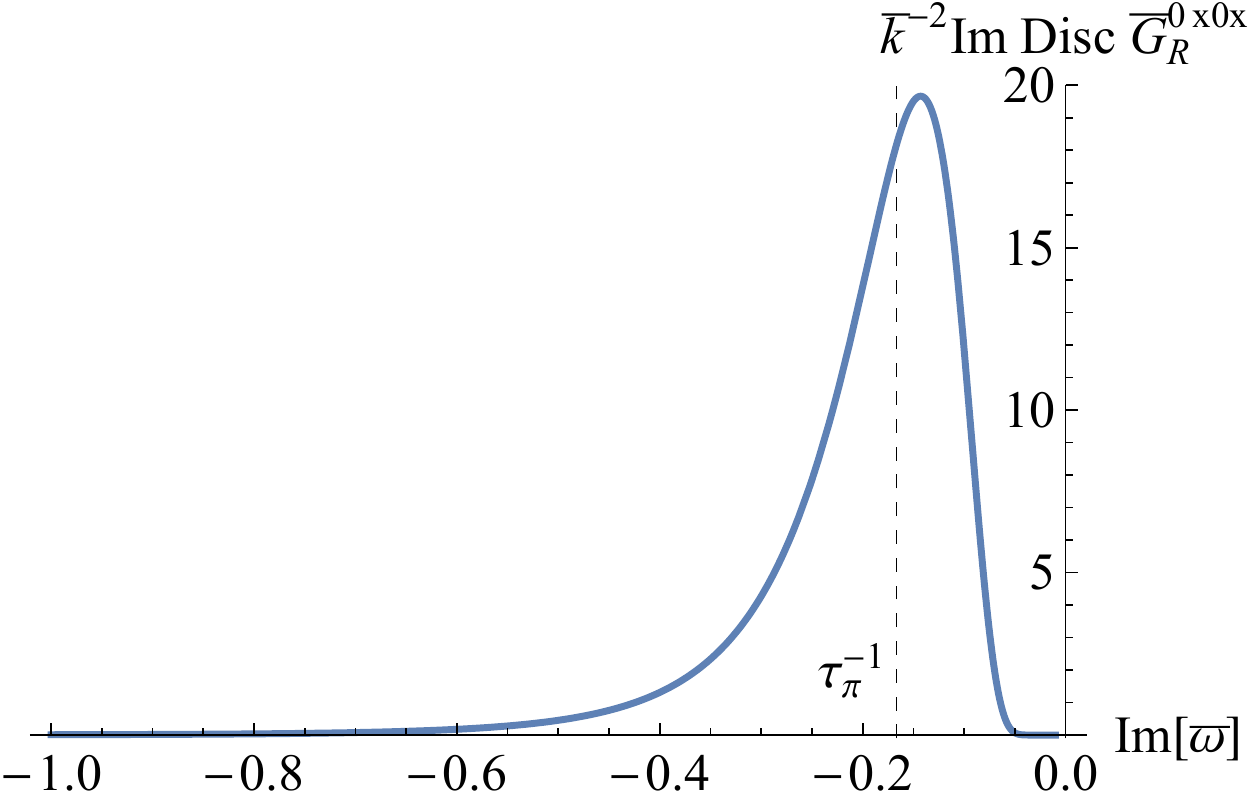}
\caption{
The suitably normalized imaginary part of the discontinuity $ \Im {\rm Disc} {\bar G}_R^{0x,0x}(\bar k + \Im(\bar\w), \bar k)/k^2 \vert_{\bar k=0}$ as a function of $\Im\bar\w$ along the cut at $\Re\bar\w = 0$. 
}
\label{fig33}
\end{figure}

According to eq.~(\ref{FT2}), the timescale over which the cut contribution dies out exponentially is inversely proportional to the depth $y$ in the complex plane
where the discontinuity becomes sizeable. The physics is particularly clear in the limit $k\to 0$, where one is dealing with one single cut and avoids issues related
to the partial cancellation between different cut contributions. In this limit, the shear viscous correlation function takes the form
\begin{align}
\frac{\bar \omega}{{\bar k}^2} {\bar G}_R^{0x0x}(\bar\w,\bar k) \vert_{\bar k = 0} = &\frac{24 \bar \omega ^5-6 i \bar \omega ^4-2 \bar\omega ^3+i \bar\omega ^2+\bar \omega}{120 \bar \omega ^6} \nonumber \\
&-\frac{i e^{i/\bar \omega } \Gamma \left(0,\frac{i}{\bar \omega }\right)}{120 \bar \omega ^6}\, .
\label{g0x0x}
\end{align}
(We note as an aside that the first nontrivial order of the shear correlator is $\propto k^2$ as a homogeneous shear perturbation corresponds to a boost and does not create shear flow.)
In Fig.~\ref{fig33}, we have plotted the suitably normalized imaginary part of the discontinuity $ {\rm Disc} {\bar G}_R^{0x,0x}(\bar k + i \bar y, \bar k)/k^2 \vert_{k=0}$ as a function of negative
$\Im(\w)$. One finds that this function peaks indeed close to $1/\tau_\pi$, thus indicating that the cut contribution to the retarded correlation function (\ref{FT2}) will be governed initially
by an exponential decay time close to $\tau_\pi$.

In summary, simple physics arguments, the numerical inspection of the imaginary part of the cut discontinuity, and the numerical calculation of the retarded correlation function shown
in Fig.~\ref{fig9} all indicate that the physical response to perturbations starts being dominated by hydrodynamics on time scales $t > \tau_\pi = 6 \td$. We emphasize, however, that it is
difficult to make this numerical observation analytically precise. The kinetic theory studied here allows for  physics on different momentum scales to relax on different time scales. 

 \subsubsection{Late time limit of the correlation function}
The late time behaviour of the correlation function is determined
by the nonanalytic structures closest to the real axis which are the cuts running to the real axis
at $\bar \w = \pm \bar k$.  In the physical response ${\bar G}_R^{\alpha \beta,\gamma \delta}(t, \bar k)$ in eq.~(\ref{FT2}),
the cut discontinuity ${\rm Disc}{\bar G}_R^{\alpha \beta,\gamma \delta}(\bar k + i \bar y, \bar k)$ at distance $\bar y= y\, \td$ from the real axis is weighted with an 
exponentially suppression $e^{{\bar y} t/\td}$. For the study of the late time behavior $t \gg 1/k$ and for sufficiently long 
wavelengths $1/k \gg \td$, i.e. $\bar k \ll 1$, it is therefore sufficient to expand this discontinuity around the
``on-shell'' point $\bar \w = \bar k$.
 
 To be specific, let us consider the shear channel correlation function ${\bar G}_R^{0x,0x}$ where the expansion of the branch cut discontinuity 
 around the on-shell point yields 
\begin{align}
{\rm Disc}{\bar G}_R^{0x,0x}(\bar k + i \bar y, \bar k) \approx -\frac{\pi  e^{1/ \bar y}}{8 \bar k {\bar y}^2 }\left( 1 + \mathcal{O}\left(\bar y \right) \right)\, .
\end{align}
The corresponding contribution to the retarded correlation function (\ref{eq58}) in the time domain reads
\begin{align}
2 & \Im e^{- i{\bar k}\, t/\td}\int_{-\infty}^{0}d y e^{{\bar y} t/\td} {\rm Disc} {\bar G}_R^{0x,0x}(\bar k + i \bar y, \bar k) \nonumber \\
&\approx - \Im  \frac{\pi  \sqrt{t/\td} e^{-i {\bar k} t/\td} K_1\left(2 \sqrt{t/\td}\right)}{2 \bar k} \nonumber \\
&\approx  \frac{\pi^{3/2} (t/\td)^{1/4}}{4 \bar k } e^{-2 \sqrt{t/\td} }\, \sin\left({\bar k}\, t/\td \right)\, .
\label{eq69}
\end{align}
This contribution to the retarded correlation function is clearly nonhydrodynamic. It is an 
oscillating function with subexponential decay, and it will therefore dominate at late times
over any contribution from hydrodynamic poles. Eq.~(\ref{eq69}) confirms in an explicit
calculation for $\xi=1$ the parametric estimates obtained for arbitrary $\xi$ in section~\ref{sec2b},
see eq.(\ref{eq17}). To estimate the scale at which this dehydrodynamization takes place,
we require that the negative exponent of the pole contribution in (\ref{FT2}) is much 
larger than the nonexponential factor in (\ref{eq69}), $t\Im(-\w_i) \gtrsim 2 \sqrt{t/\td}$. 
Since the imaginary part of the fluid dynamic poles $\w_i$ starts $\propto k^2$ for small $k$, we therefore conclude that in 
the scale-dependent relaxation time approximation
investigated here, the kinetic theory dehydrodynamizes for arbitrarily small $k$ at sufficiently late
times, 
\begin{align}
	t \gtrsim \frac{1}{\Im\left[-\w_i(k)\right]^2\, \td}\, .
	\label{eq74}
\end{align}
This dehydrodynamization is visible in an oscillatory sub-exponential
late-time decay of retarded correlation functions, as can be seen in Fig.~\ref{fig9}.
   
\begin{figure}
\includegraphics[width=0.5\textwidth]{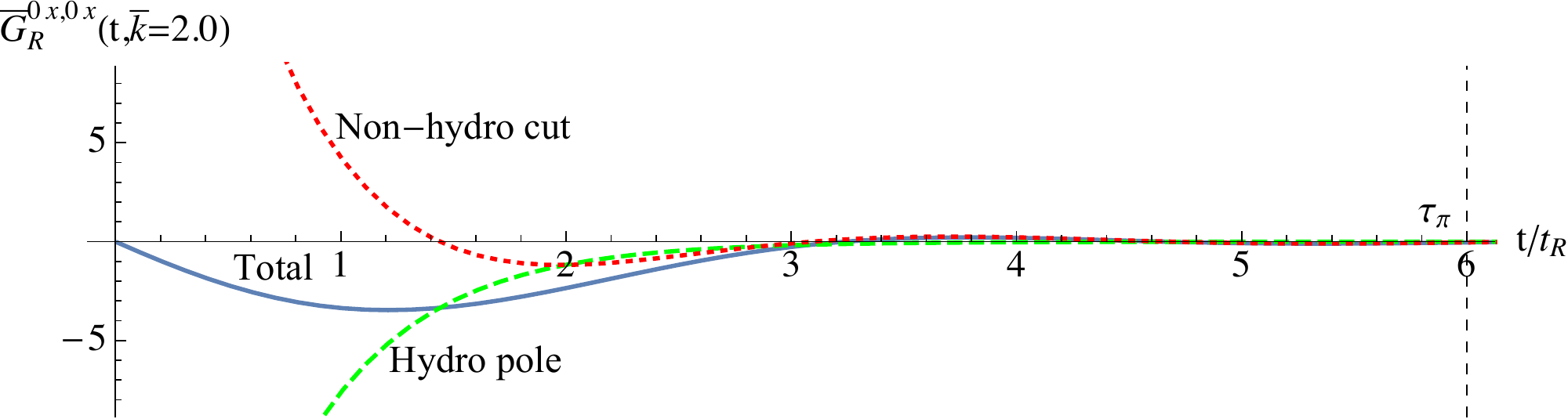}
\caption{
Same as Fig.~\ref{fig9}, but now for a larger momentum scale $\bar k = 2$ for which the hydrodynamic pole does not dominate the time evolution on any timescale.
}
\label{figpp}
\end{figure}

According to eq.(\ref{eq74}), the timescale at which dehydrodynamization occurs  varies strongly with the momentum $k$. While Fig.~\ref{fig9} shows a wide window of close-to-hydrodynamic
evolution for $\bar k = 0.4$, this window closes if $\bar k $ is increased to values larger than unity. As seen in Fig.~\ref{figpp}, already for ${\bar k} = 2$, the oscillatory late-time behavior is visible
at all time-scales and a window of close-to-hydrodynamic behavior does not exist.

\section{Asymptotic nature of gradient expansion and Borel summability}
\label{sec6}
\begin{figure}
\includegraphics[width=.45 \textwidth]{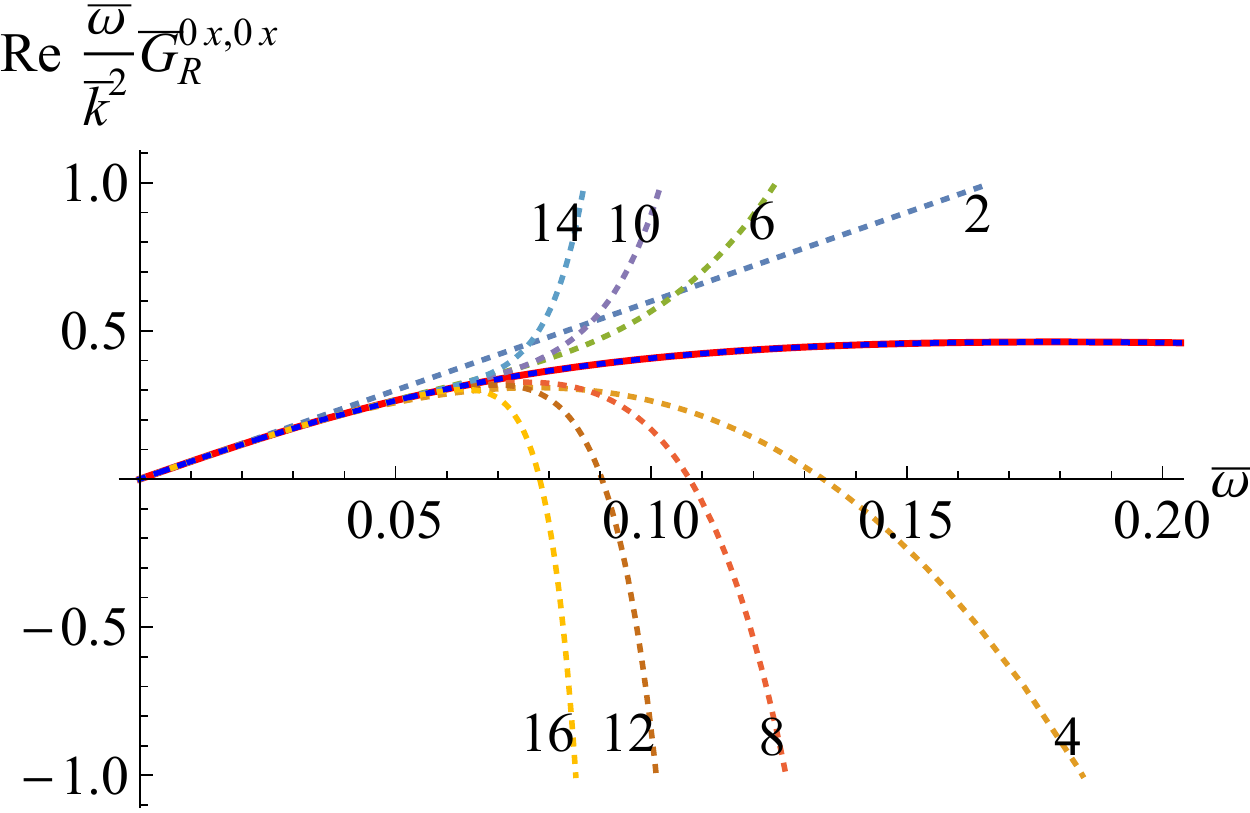}
\caption{ The real part of the shear channel correlation function (\ref{g0x0x}) (red line) compared to gradient expansions (\ref{eq79}).
The integers $N$ on the dashed lines denote the highest power $\propto \bar\w^{N}$ included
in the gradient expansion. 
 The blue dashed line on top of the red line is the 25th order Pad\'e approximant.}
\label{Pade}
\end{figure}

We discuss now the use of Borel techniques to resum the divergent gradient series of the correlation functions. To simplify the discussion and to arrive at analytical expressions, we consider 
the shear channel correlation function (\ref{g0x0x}) in the limit of vanishing $k$ as an explicit example.
Its hydrodynamical gradient expansion corresponds to a Taylor expansion in $\omega$
\begin{align}
\frac{\bar \omega}{\bar k^2} {\bar G}_R^{0x0x}(\bar \w,\bar k)\vert_{\bar k = 0} \approx \sum_{ i=0} ^{N} b_j {\bar \w}^j\, .
\label{eq79}
\end{align}
Comparing the full expression to different orders of this gradient expansion, one sees from Fig.~\ref{Pade} that the expansion in powers of $\omega$ is an asymptotic. For a given value of $\bar \w$, inclusion of higher order terms does not improve the approximation but instead makes it worse. This poor convergence of the series is caused by a factorial growth in the Taylor coefficients and is a consequence of the cut of the $\Gamma$-function extending to the expansion point $\bar \omega = 0$. 

A standard trick for improving the convergence of the series near non-analytic structures is to replace the Taylor series by a Pad\'e approximant
\begin{align}
\frac{\bar \omega}{\bar k^2} {\bar G}_R^{0x0x}(\bar \w,\bar k)\vert_{\bar k = 0} \approx \frac{\sum_{i} c_i \bar \omega^i}{\sum_{j} d_j \bar \omega^j}\, ,
\end{align}
 which as a rational polynomial can account for non-analytic structures. Indeed, as is evident from Fig.~\ref{Pade}, the 25th order Pad\'e approximant (that is, approximating the function with rational polynomial whose numerator and denominator are 25th order polynomials in $\omega$, and whose Taylor expansion coincides with that of the original function up the $\bar \omega^{50}$) performs vastly better numerically. 

\begin{figure}
\includegraphics[width=0.45\textwidth]{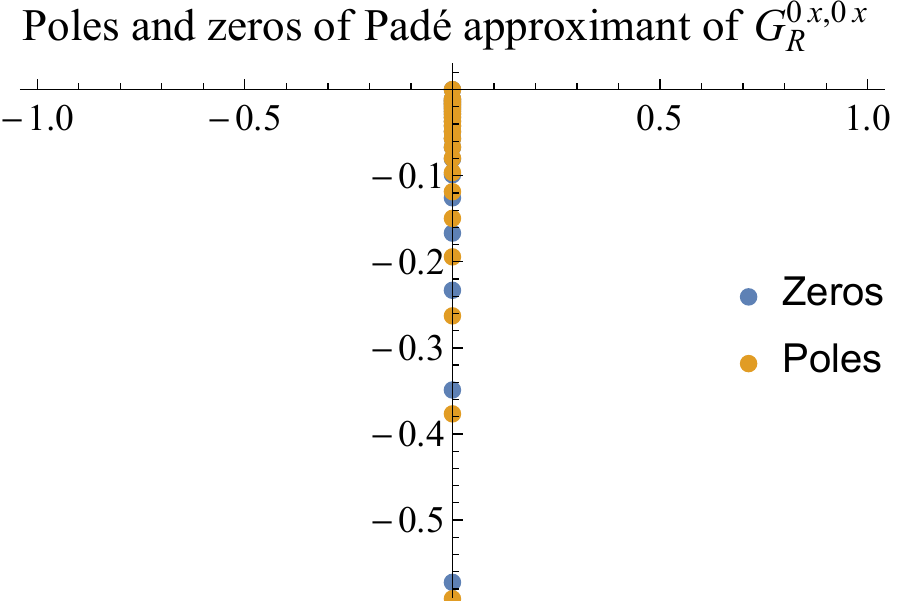}
\caption{Poles and zeros of 25th order Pad\'e approximant of the shear channel correlation function. The cut of the $\Gamma$-function is mimicked by a string of poles and zeros. }
\label{fig15}
\end{figure}

Whereas the non-analytic structure of the correlation function is a cut, the only non-analytic structures present in the Pad\'e approximant are poles. The way the cut is mimicked by the Pad\'e approximant is in term of an alternating string of poles and zeroes where the original cut lies, such that the poles become denser as the order of approximation is increased, see Fig.~\ref{fig15}. 

In order to gain further improvement, one may try to use Borel's trick of writing factorials in integral representation, $j! = \int_0^\infty s^{j}\, e^{-s}\, ds$,
  \begin{align}
\frac{\bar \omega}{\bar k^2} {\bar G}_R^{0x0x}(\bar \w,\bar k)\vert_{\bar k = 0}  &= \sum_{j=1}^n b_j\,{\bar \w}^{j} \nonumber\\
& = \int_0^\infty e^{-s} \left(\sum_{j=1}^\infty \frac{b_j}{j!}\,  (s{\bar \w})^{j} \right)\, .
 \label{eqc1}
 \end{align}
 The art is then to perform the Borel sum in the integrand of (\ref{eqc1}) which can be convergent since it has factorially suppressed coefficients.
As typically one has information only of finite set of Taylor coefficients $b_j$, the standard practice is to again approximate the Borel transform 
\begin{align}
B(s) = \sum_{j=1}^\infty \frac{b_j}{j!}\,  s^{j}
\end{align}
using a Pad\'e approximant. 

In our case it turns out that the Borel transform is itself a rational function, and therefore the Pad\'e approximation is exact once a required amount of terms are taken into account  
\begin{align}
B(s) = \frac{i}{(i + s)^6}\, .
\end{align}
Of course, if we had access only to a finite number of Taylor coefficients, we could not know for sure that we have fully reconstructed the Borel transform. But in our case, we may simply compute the inverse transformation of eq.~(\ref{eqc1}), and indeed we recover back the original expression (\ref{g0x0x}). It is remarkable how using the Borel resummation we have been able to recover the non-analytic features of the correlation function with only perturbative information about the gradient series. 

It has been suggested that the non-analytic features in the Borel transform arise from physics of non-hydrodynamical modes. In the example at hand, it is easy to see that the essential singularity at the origin arises from the residue of the only pole of the Borel transform
\begin{align}
\oint ds e^{-s}\frac{i}{(i + s )^6} = -\frac{e^{i/\bar \w}}{240 \pi  \bar \w^6}\, .
\end{align}
We find it curious that the exponent in the previous equation, or the location of the 
nonanalyticity of the Borel transformation are not directly related to the location of the nonhydrodynamic mode with the smallest imaginary part. This is in contrast to the analogous problem in an expanding background, where the system is driven out of equilibrium because of longitudinal expansion instead of a external metric source. It has been suggested in~\cite{Heller:2013fn,Heller:2015dha,Heller:2016rtz}
 that in this case
the location of the first nonanalytic structure in the Borel plane is given by the slowest decaying nonhydrodynamic mode.
In our case, the nonhydrodynamic mode with the smallest imaginary part has always vanishing imaginary part, and indeed the nonanalytic behaviour arises from the combined effect of all 
nonhydrodynamic modes. 

To contrast this picture with a case where the nonhydrodynamic modes are well separated from the expansion point of $\bar \omega=0$, consider the correspoding shear channel correlation function at vanishing $\bar k$ in the case of $\xi = 0$
\begin{align}
\frac{\bar \omega}{\bar k^2} {\bar G}_R^{0x0x}(\bar \w,{\bar k})\vert_{\bar k = 0} = \frac{1}{5}\frac{1}{\bar \w + i}.
\end{align}
In this trivial case the gradient expansion is well behaved and the pole located at $\bar \w = -i$ sets the radius of convergence. In this case the Borel transformation reads 
\begin{align}
B(s) = \frac{e^{- i s}}{5}\, ,
\end{align}
which is a complete function with only an essential singularity at large $s$.

We also note that, from the point of view of Borel summation, the cases $xi=0$ and $\xi=1$ are both speacial. For $\xi = 0$, the gradient expansion is convergent series. For $0 < \xi \leq 1$, its 
Borel sum is convergent while the gradient expansion itself is asymptotic. As seen, e.g., from eq.~(\ref{eq44b}),
the coefficients of the gradient expansion for $\xi > 1$ grow faster than factorial, making also the Borel sum nonconvergent.

\section{Conclusions}
Generically, the path to equilibration in relativistic systems described by Boltzmann transport is governed by an interplay of collective hydrodynamic and non-collective particle
excitations. The present study allowed us to expose this interplay in detail. Generically, there is no sharp onset of hydrodynamic behavior. On all time 
and length scales, both hydrodynamic and non-hydrodynamic modes are present. To which extent the one dominates over the other can be at best a quantitative
statement that changes gradually with scale. Also, the appearance of poles in the first (physical) Riemann sheet of retarded correlation functions is a
matter of choosing a particular analytical continuation and thus cannot be related unambiguously to the onset of fluid dynamic behavior. 
Still, even if the pole can be made disappear from the physical Riemann sheet by utilizing the ambiguity in analytic continuation, its weight is translated unambiguously
to other non-analytic structures in that sheet. In this sense, the relative closeness of hydrodynamic poles to the real axis carries quantitative information about the onset of hydrodynamic behavior
irrespective of whether they are visible.

The hydrodynamic behavior is fully characterized by the coefficients of a gradient expansion. As we showed for a generic kinetic theory,  this expansion is asymptotic already
for retarded correlation functions, since the starting point of the branch cut approaches the origin for small $k$. This is in marked different to results obtained for strong coupled
field theories and in the standard scale-independent relaxation time approximation, where the gradient expansion for retarded correlation functions converges. Remarkably,
however, the latter theories if pushed out of equilibrium by longitudinal expansion exhibit a time-dependent  energy density whose gradient expanion (in powers of inverse time) 
is asymptotic. We note that non-linear transport coefficients appear in this expansion, while the above-mentioned gradient expansions of retarded correlation functions involve
linear transport coefficients only. It would be interesting to understand the relation between the analytic structures of the higher $n$-point functions that give rise to non-linear
transport coefficients, and the qualitatively different convergence properties of the above-mentioned gradient expansions. 

Borel summation is employed in attempts to extract physically meaningful information from non-convergent asymptotic series. This technique is often advocated with the
seemingly contradictory claim that it can reveal non-perturbative information from analysis of purely perturbative input. 
By explicitly resumming the Borel series of the gradient expansion of a retarded correlator, we demonstrated in section~\ref{sec6} how this can function.
To the best of our knowledge, this is the first time that an explicit Borel transformation of a hydrodynamizing non-equilibrium system has been  fully performed.  

Our study could be extended on several fronts. The present discussion remained limited to linear response and it could be extend within the present set-up to non-linear
response and, in line with the remarks above, to systems undergoing expansion. It may also be interesting to supplement the kinetic theories studied here with thermal
fluctuations that via the fluctuation-dissipation theorem are known to give rise to characteristic long-time hydrodynamical tails. Furthermore, it would be interesting to 
observe, e.g., in numerical simulations, the features identified here in kinetic theories whose collision kernels are derived directly from quantum field theory. 
Finally, as mentioned in the introduction, a full quantum field theoretical treatment contains interference effects that  go beyond simple kinetic theory and become
relevant at higher orders in perturbation theory.

\appendix
\section{Calculation of the integral moments $I^{abc}$ for $\xi = 1$}
\label{appb}
In this appendix, we provide further information on how to evaluate the integral moments $I^{abc}$ of eq.~(\ref{eq30}), which can be written for $\xi = 1$ in the form
\begin{align}
	I^{a,b,c} = &
	\frac{1}{24}  \int dp\, p^{5-a}\, e^{-p}\, 
	\int \frac{d\phi}{2\pi} \sin^b\phi \nonumber \\
	& \times
	\int_{-1}^1 \frac{dx}{2} \frac{\left(1-x^2\right)^{(b/2)}\, x^c}{1 + p\, (-i\bar{\omega} + i \bar{k}x )}\, .
	\label{eqb1}
\end{align}
Here, $\bar \w \equiv \td \w$, $\bar k \equiv \td k$ and $x$ denotes the cosine of the angle between $\vec{v}$ and $\vec k$.
The $\phi$-integration leads to trivial prefactors. Only integral moments with even integer index $b$ are non-vanishing. 
To bring the $p$- and $x$-integrations into a simpler form, we proceed as follows: We first observe that 
for $b=c=0$, the elementary $x$-integral returns a logarithm
\begin{align}
	\int_{-1}^1 \frac{dx}{2} \frac{p}{1 + p (-i\bar{\omega} + i \bar{k}x )} &= -\frac{i}{2 \bar{k}} \log\left[ \frac{i-p\, \bar{k} + p \bar{\w}}{i+p\, \bar{k} + p\, \bar{\w}} \right] \nonumber \\
	& \equiv -\frac{i}{2 \bar{k}} L\, .
	\label{eqb2}
\end{align}
For arbitrary positive integers $b$, $c$, the corresponding integral can be shown to be of the form
\begin{align}
	&\int_{-1}^1 \frac{dx}{2} \frac{p\, \left(1-x^2\right)^{(b/2)}\, x^c}{1 + p (-i\bar{\omega} + i \bar{k}x )} \int \frac{d\phi}{2\pi} \sin^b\phi  \nonumber \\
	&= T^{b,c}_1(\bar{k}\, p,\bar{\w}\, p)+ T^{b,c}_2(\bar{k}\, p,\bar{\w}\, p)  \frac{-i}{2 \bar{k}} L \, .
	\label{eqb3}
\end{align}
For the components relevant for our calculation, we have tabulated the functions $T^{b,c}_1(\bar{k} p,\bar{\w} p)$ and $T^{b,c}_2(\bar{k} p,\bar{\w}p)$ 
in Table~\ref{table1}. To obtain the moments $I^{a,b,c}$ in (\ref{eqb1}), it then remains to perform the integral 
\begin{align}
	&I^{a,b,c}(\bar k,\bar \w) \nonumber\\
	 & = \frac{1}{24} \int dp\, p^{4-a}\, e^{-p}\, 
	\left( T_1^{b,c}(\bar{k}p,\bar{\w}p)+ T_2^{b,c}(\bar{k}p,\bar{\w}p)  \frac{-i}{2 \bar{k}} L  \right)\, . 
	\label{eqb4}
\end{align}
For all moments that enter the retarded correlation functions (\ref{tensor}), (\ref{shear}) and (\ref{sound}), 
the products $p^{4-a}\, T_1^{b,c}(\bar{k}p,\bar{\w}p)$ and $p^{4-a}\, T_2^{b,c}(\bar{k}p,\bar{\w}p)$ in the integrand of eq.(\ref{eqb4}) 
are explicitly known polynomials in $p$ that include only positive powers up to $p^4$.
The first term in (\ref{eqb4}) is then easily integrated, using  
\begin{align}
\int dp\, p^n\, e^{-p} = \Gamma[n+1]\, .
\label{eqb5}
\end{align} 
The second term in (\ref{eqb4}) requires calculating for $n=0,1,2,3,4$ the expression
\begin{align}
	&\frac{1}{24} \int dp\, p^{n}\, e^{-p}\,  \frac{-i}{2 \bar{k}} L \nonumber\\
	&= \frac{i}{24} \int dp\, p^{n}\, e^{-p}\, \int_{-1}^1 \frac{dx}{2} \frac{p}{(\bar{\omega} - \bar{k}x )p + i } \nonumber \\
	& = \frac{i}{24} \left(-1\right)^n \partial_{\rho}^{n} \int_{-1}^1 \frac{dx}{2} \int_0^\infty dp\, 
	\frac{p\, e^{-\rho\, p}}{(\bar{\omega} - \bar{k}x )p + i } \Big\vert_{\rho = 1} \nonumber \\
	& = \frac{i}{24} \left(-1\right)^n \partial_{\rho}^{n}\, H(\bar \w,\bar k,\rho)  \Big\vert_{\rho = 1}\, ,
	\label{eqb6}
\end{align}
where $H(\bar \w,\bar k,\rho)$ is the analytically known generating function defined in  (\ref{eq49}), or an analytically continued function $H_a$ that agrees with
$H$ along the real axis and the positive imaginary $\bar \w$-half plane.

In this way, somewhat lengthy but explicit expressions for all relevant integral moments $I^{a,b,c}(\bar \w, \bar k)$  are obtained by inserting into (\ref{eqb4})  the explicit terms given  
in table~\ref{table1}, writing these terms in powers of $p$, and performing the $p$-integrals with the help of eqs.(\ref{eqb5}) and (\ref{eqb6}).
\begin{table}[h]
\begin{center}
    \begin{tabular}{|  c | c | c | c | }
    \hline
    $b$ & $c$ & $T^{b,c}_1(k,\w)$ & $T^{b,c}_2(k,\w)$ \\ \hline
    0 & 0 & $0$ & $1$ \\ \hline
    0 & 1 & $\textstyle\frac{-i}{ k}$ & $\textstyle\frac{i+\w}{k}$ \\ \hline
    0 & 2 & $\textstyle\frac{1-i \w}{{k}^2}$ & $\textstyle\frac{(i+\w)^2}{{ k}^2}$ \\ \hline
    0 & 3 & $\textstyle\frac{-i \left(k^2 + 3 (i+\w)^2\right)}{3\, k^3}$ & $\textstyle\frac{(i+\w)^3}{{ k}^3}$  \\ \hline
    0 & 4 & $\textstyle\frac{\left(1-i\w \right) \left(k^2 + 3 (i+\w)^2\right)}{3\, k^4}$ & $\textstyle\frac{(i+\w)^4}{{ k}^4}$  \\ \hline
    2 & 0 & $\textstyle\frac{1}{2}\textstyle\frac{- \left(1-i \w\right)}{{k}^2}$ & $\textstyle\frac{1}{2}\textstyle\frac{k^2 - (i+\w)^2}{{ k}^2}$ \\ \hline
    2 & 1 & $\textstyle\frac{1}{2}\textstyle\frac{-i \left(2 k^2 - 3 (i+\w)^2\right)}{3\, k^3}$ & $\textstyle\frac{1}{2}\textstyle\frac{(i+\w)\, \left( k^2 - (i+\w)^2 \right)}{{ k}^3}$ \\ \hline
    2 & 2 & $\textstyle\frac{1}{2}\textstyle\frac{\left(1-i\w \right) \left(2 k^2 - 3 (i+\w)^2\right)}{3\, k^4}$ & $\textstyle\frac{1}{2}\textstyle\frac{(i+\w)^2\, \left( k^2 - (i+\w)^2 \right)}{{ k}^3}$ \\ \hline
    4 & 0 & $\textstyle\frac{3}{8}\textstyle\frac{\left(1-i\w \right) \left(-5 k^2 + 3 (i+\w)^2\right)}{3\, k^4}$ & $\textstyle\frac{3}{8}\textstyle\frac{\left( k^2 - (i+\w)^2\right)^2}{{ k}^4}$ \\
    \hline
    \end{tabular}
\end{center}
\caption{The functions $T^{b,c}_1(k,\w)$ and $T^{b,c}_2(k,\w)$ that appear in eq.~(\ref{eqb3}).}
\label{table1}
\end{table}

\acknowledgements
We thank the organizers and participants of "Micro-workshop on analytic properties of thermal correlators at weak \& strong coupling" at Oxford in March 2017 for giving the initial inspiration to this work. We thank Peter Arnold, Harvey Meyer, Krishna Rajagopal, and Andrei Starinets for useful discussions.

\end{document}